\documentclass[letterpaper,british, aps,prb,floatfix,twocolumn,showpacs,amsmath,amssymb,eqsecnum,superscriptaddress]{revtex4-1}
\usepackage[T1]{fontenc}
\usepackage[latin9]{inputenc}
\usepackage{color}
\usepackage{mathrsfs}
\usepackage{amsmath}
\usepackage{amssymb}
\usepackage{graphicx}

\makeatletter

\pdfpageheight\paperheight
\pdfpagewidth\paperwidth

\providecommand{\tabularnewline}{\\}
\newcommand{\lyxdot}{.}

\newcommand{\angstrom}{\text{\normalfont\AA}}
\usepackage{hyperref}

\makeatother

\usepackage{babel}
\begin{document}

\preprint{APS/123-QED}

\title{Many-body renormalisation of forces in $f$-electron materials}

\author{Evgeny Plekhanov}
\email{evgeny.plekhanov@kcl.ac.uk}

\selectlanguage{british}%

\affiliation{King's College London, Theory and Simulation of Condensed Matter
(TSCM), The Strand, London WC2R 2LS, UK}

\author{Phil Hasnip}

\affiliation{Department of Physics, University of York, Heslington, York YO10
5DD, UK}

\author{Vincent Sacksteder}

\affiliation{Department of Physics, Royal Holloway University of London, Egham,
Surrey TW20 0EX, UK}

\author{Matt Probert}

\affiliation{Department of Physics, University of York, Heslington, York YO10
5DD, UK}

\author{Stewart J. Clark}

\affiliation{Department of Physics, University of Durham, Durham DH1 3LE, UK}

\author{Keith Refson}

\affiliation{ISIS Facility, RAL, Chilton, Didcot OX11 0QX, UK}

\affiliation{Department of Physics, Royal Holloway University of London, Egham,
Surrey TW20 0EX, UK}

\author{Cedric Weber}

\affiliation{King's College London, Theory and Simulation of Condensed Matter
(TSCM), The Strand, London WC2R 2LS, UK}

\date{\today}
\begin{abstract}
We present the implementation of Dynamical Mean-Field Theory (DMFT)
in the CASTEP \emph{ab-initio} code.
We explain in detail the theoretical
framework for DFT+DMFT and we demonstrate our implementation for three
strongly-correlated systems with $f$-shell electrons: $\gamma$-cerium,
cerium sesquioxide Ce$_{2}$O$_{3}$ and samarium telluride SmTe by using a Hubbard I solver.
We find very good agreement with previous benchmark DFT+DMFT calculations of cerium compounds, while
for SmTe, which was never studied within DFT+DMFT before to the best of our knowledge, we show the improved agreement with the
experimental structural parameters as compared with LDA.
Our implementation works equally well for both
norm-conserving and ultra-soft pseudopotentials, and we apply it to
the calculation of total energy, bulk modulus, equilibrium volumes
and internal forces in the two cerium compounds. In Ce$_{2}$O$_{3}$
we report a dramatic reduction of the internal forces acting on coordinates
not constrained by unit cell symmetries. This reduction is induced
by the many-body effects, which can only be captured at the DMFT level.
In addition, we derive an alternative form for treating the high-frequency
tails of the Green function in Matsubara frequency summations. Our
treatment allows a reduction in the bias when calculating the correlation
energies and occupation matrices to high precision. 
\end{abstract}



\maketitle

\section{INTRODUCTION}

Density functional theory (DFT) is a key computational tool for modern
material science, condensed matter physics and solid-state chemistry\citep{DFT-HK,DFT-KS,Jones_2015}.
It can treat an immense range of materials, including bulk metals,
oxides, semiconductors, graphene and layered materials, and surfaces.
Practical DFT calculations, however, rely on approximate exchange-correlation
functionals, which handicaps the ability of DFT to reproduce strongly
correlated physics in many materials, notably those containing open
$d$ or $f$-shell elements. Many strongly-correlated materials exhibit
properties useful for technological applications\citep{Kotliar_Vollhardt_2004,Weber_2012,Weber_2013}.
For example, the copper oxides and iron pnictides are high temperature
superconductors\citep{Plekhanov_2003,Plekhanov_2005,Dai_2015}, and
the cobaltates exhibit colossal thermoelectric power\citep{Lija_2014}
which is useful for energy conversion. Several vanadates have peculiar
room-temperature metal-insulator transitions, allowing realisation
of a so-called \textquotedblleft intelligent window\textquotedblright ,
which becomes insulating as the external temperature 
drops\citep{Babulanam_1987,Granqvist_1990,Granqvist_2007,Tomczak_2009}.
The failure of DFT's exchange-correlation functionals to capture strong
correlation physics severely limits its use for nano-scale design
of these many, important functional materials.

In contrast to DFT, huge progress has been made
in describing strongly-correlated materials with Dynamical Mean-Field
Theory (DMFT)\citep{Georges_1996,Vollhardt_2010,Savrasov_2004,Minar_2005,Kotliar_2006,Pourovskii_2007}.
DMFT is a sophisticated method which offers a higher level of theoretical
description than DFT, and bridges the gap between DFT and Green function
approaches. Within DMFT, the treatment of local electronic correlation
effects is formally exact, although the non-local electronic correlation
effects are neglected.

In this work, we provide a fast and stable implementation of the full
charge self-consistent DFT+DMFT moethod in the widely used plane-wave
DFT code CASTEP\citep{DFT-HK,DFT-KS,RMP-Payne,CASTEP},
and benchmark this implementation by calculating spectral properties,
energetics and forces for $\gamma$-Ce, Ce$_{2}$O$_{3}$ and SmTe. It was
shown previously\citep{Pourovskii_2007}, that full charge self-consistency
is not crucial for these compounds and the Hubbard I solver (at least
at the level of total energy). Therefore, in this manuscript, we focus
on the DMFT approach within the framework of fixed Kohn-Sham (KS)
potentials, the so-called ``one-shot'' DFT+DMFT method. We show
that our predicted equilibrium volume and bulk modulus for cerium
compounds are in excellent agreement with the existing literature,
\emph{i.e.} that taking into account strong correlations improves
the agreement with the experiment compared to DFT. Moreover, by calculating
the atomic forces in cerium sesquioxide we show that DFT overestimates
them by almost a factor of two.

The rest of this paper is organised as follows: in Section \ref{sec:Methods},
we re-derive the DFT+DMFT formalism in the case of plane-wave basis;
in Sections \ref{sec:Examples}-\ref{sec:Forces}, we illustrate our
results on the examples of $\gamma$-Ce and cerium sesquioxide; finally
Section \ref{sec:Conclusions} is dedicated to the conclusions.

\section{\label{sec:Methods}Methods}

\subsection{General formalism}

There exist in the literature several proposals for combining DFT
and DMFT\citep{Pourovskii_2007,Amadon_2008,Lichtenstein2013,Lechermann_2006,BENEDICT_1993}. Here,
we follow closely the DFT+DMFT formulation proposed in Refs.\onlinecite{Pourovskii_2007,Amadon_2008}.
Nevertheless, in contrast with the Ref.\onlinecite{Pourovskii_2007},
where an LMTO basis was considered, we deal with a plane-wave code
CASTEP. On the other hand, contrarily, to Ref.\onlinecite{Amadon_2008},
we use a different orthogonalisation procedure. We therefore, re-derive
all the formulae, relevant for our case taking into account these
differences.

The total energy functional was derived in Refs.\onlinecite{Savrasov_2004,Kotliar_2006,Pourovskii_2007}
and is reported here for completeness. The starting point is the Baym-Kadanoff
(or Luttinger-Ward) functional (for a review see Ref.\onlinecite{Kotliar_2006}),
which is a functional of electronic density $\rho(\mathbf{r})$ and
an impurity Green function $G_{m,m^{\prime}}^{\mathbf{R}}\left(i\omega_{n}\right)$
with the respective constraint fields $v_{KS}(\mathbf{r})$ and $\Sigma_{m,m^{\prime}}^{\mathbf{R}}\left(i\omega_{n}\right)$:
\begin{align}
 & \Omega\left[\rho,G_{m,m^{\prime}}|v_{KS},\Sigma_{m,m^{\prime}}\right]=\mathrm{Tr}\ln\hat{G}\nonumber \\
 & -\int d\mathbf{r}\left(v_{KS}(\mathbf{r})-v_{c}(\mathbf{r})\right)\rho(\mathbf{r})-\mathrm{Tr}G\Sigma\nonumber \\
\label{eq:Baym}\\
 & +\frac{1}{2}\int d\mathbf{r}d\mathbf{r}^{\prime}\rho(\mathbf{r}^{\phantom{\prime}})\frac{1}{\left|\mathbf{r}-\mathbf{r}^{\prime}\right|}\rho\left(\mathbf{r}^{\prime}\right)+E_{xc}[\rho]\nonumber \\
 & +\sum_{\mathbf{R}}\left(\Phi_{imp}[G_{m,m^{\prime}}^{\mathbf{R}}]-\Phi_{DC}[G_{m,m^{\prime}}^{\mathbf{R}}]\right).\nonumber 
\end{align}
Here, $G_{m,m^{\prime}}^{\mathbf{R}}\left(i\omega_{n}\right)$ and
$\Sigma_{m,m^{\prime}}^{\mathbf{R}}\left(i\omega_{n}\right)$ are
defined as matrices in orbital indices $m$ and $m^{\prime}$ and
functions of Matsubara frequencies $i\omega_{n}$, $E_{xc}[\rho]$
is the exchange-correlation functional, $v_{c}(\mathbf{r})$ is the
periodic potential of the ions, $\Phi_{imp}[G_{m,m^{\prime}}^{\mathbf{R}}]$
is the DMFT interaction functional and $\Phi_{DC}[G_{m,m^{\prime}}^{\mathbf{R}}]$
is the double-counting functional. Finally, $\hat{G}$ is the Bloch
Green function operator:
\begin{equation}
   \hat{G}(\mathbf{r},i\omega_{n})=\left(i\omega_{n}+\mu+\frac{1}{2}\nabla^{2}
   -v_{KS}(\mathbf{r})-\Sigma^{B}(\mathbf{r},i\omega_{n})\right)^{-1}.\label{eq:GF}
\end{equation}
$\Sigma^{B}(\mathbf{r},i\omega_{n})$ is the Bloch self-energy obtained
by up-folding of $\Sigma_{m,m^{\prime}}^{\mathbf{R}}$ (explained
below), while $\mathrm{Tr}A$ of a matrix function (or operator) is
the shorthand notation for:
\begin{equation}
\mathrm{Tr}A=T\sum_{n,l}A_{ll}(i\omega_{n})e^{i\omega_{n}0^{+}},
\end{equation}
\emph{i.e.} traced over both orbital and imaginary time indices at
temperature $T$. Here, we use the Atomic Hartree units, so that $\hbar=1$,
$e=1$ and $m_{e}=1$. The variation of $\Omega$ with respect to
$\rho$ and $G_{m,m^{\prime}}^{\mathbf{R}}$ gives the constraint
potentials $v_{KS}$ and $\Sigma_{m,m^{\prime}}^{\mathbf{R}}$ respectively:
\begin{align}
v_{KS}(\mathbf{r}) & =v_{c}(\mathbf{r})+\frac{\delta E_{xc}}{\delta\rho}+\int d\mathbf{r}^{\prime}\frac{1}{\left|\mathbf{r}-\mathbf{r}^{\prime}\right|}\rho\left(\mathbf{r}^{\prime}\right)\nonumber \\
\label{eq:constraints}\\
\Sigma_{m,m^{\prime}}^{\mathbf{R}} & =\frac{\delta\Phi_{imp}}{\delta G_{m,m^{\prime}}^{\mathbf{R}}}-V^{DC}.\nonumber 
\end{align}
Here $V_{DC}$ is the double counting potential:
\begin{equation}
V^{DC}=\frac{\delta\Phi_{DC}[G_{m,m^{\prime}}^{\mathbf{R}}]}{\delta G_{m,m^{\prime}}^{\mathbf{R}}},
\end{equation}
 while the variation of $\Phi_{imp}$ with respect to $G_{m,m^{\prime}}^{\mathbf{R}}$
is by construction the outcome of the impurity solver \textendash{}
the impurity self-energy:
\begin{equation}
\frac{\delta\Phi_{imp}[G_{m,m^{\prime}}^{\mathbf{R}}]}{\delta G_{m,m^{\prime}}^{\mathbf{R}}}=\Sigma_{m,m^{\prime}}^{imp}.
\end{equation}
 On the other hand, the variation with respect to $v_{KS}$ and $\Sigma_{m,m^{\prime}}^{\mathbf{R}}$,
taking into account \eqref{eq:constraints} yields $\rho$ and $G_{m,m^{\prime}}^{\mathbf{R}}$
respectively:
\begin{align}
\rho\left(\mathbf{r}\right) & =\mathrm{Tr}\langle\mathbf{r}\left|\hat{G}\right|\mathbf{r}\rangle\nonumber \\
\label{eq:min_rho}\\
G_{m,m^{\prime}}^{\mathbf{R}} & =\left\langle \chi_{m\mathbf{R}}\left|\hat{G}\right|\chi_{m^{\prime}\mathbf{0}}\right\rangle ,\nonumber 
\end{align}
where $\left\{ \chi_{m\mathbf{R}}\right\} $ is the localised basis,
used to define the Coulomb interaction. Here indices $m\mathbf{R}$
signify $m$-th orbital on ion sitting at position $\mathbf{R}$.
We will also use in what follows an abbreviated notation including
spin notation $\sigma$: $\left\{ m\mathbf{R}\sigma\right\} =L$.
From \eqref{eq:constraints}, the constraint field $v_{KS}$ and $\Sigma_{m,m^{\prime}}$
can be expressed in terms of $\rho$ and $G_{m,m^{\prime}}^{\mathbf{R}}$.
We thus arrive at the functional $\Gamma$, which is a functional
of only $\rho$ and $G_{m,m^{\prime}}^{\mathbf{R}}$:
\begin{equation}
\Gamma[\rho,G_{m,m^{\prime}}^{\mathbf{R}}]=\Omega\left[\rho,G_{m,m^{\prime}}^{\mathbf{R}}|v_{KS}[\rho],\Sigma_{m,m^{\prime}}^{\mathbf{R}}\left[G_{m,m^{\prime}}^{\mathbf{R}}\right]\right].
\end{equation}
 Finally, the minimum free-energy is obtained by noting that at minimum\citep{Kotliar_2006}
$\Gamma[\rho,G_{m,m^{\prime}}^{\mathbf{R}}]=F[\rho,G_{m,m^{\prime}}^{\mathbf{R}}]$.
Thus, substituting $\rho$ and $G_{m,m^{\prime}}^{\mathbf{R}}$ and
\eqref{eq:constraints} into \eqref{eq:Baym} gives the minimal value
of the free energy. At zero temperature, the free energy reduces to
the total (internal) energy, which can be rewritten using the DFT
total energy\citep{Pourovskii_2007}:

\begin{align}
E_{tot} & =E_{DFT}-\sum_{\nu,\mathbf{k}}f_{\nu}^{DFT}(\mathbf{k})\varepsilon_{\mathbf{k},\nu}^{DFT}\label{eq:E_tot}\\
 & +\sum_{\nu,\mathbf{k}}N_{\nu,\nu}(\mathbf{k})\varepsilon_{\mathbf{k},\nu}+E_{U}-E^{DC}.\nonumber 
\end{align}
Here $k$ is the crystal momentum, $\nu$ is the band index, $E_{DFT}$
is the total energy of underlying DFT calculations, $f_{\nu}^{DFT}(\mathbf{k})$
and $N_{\nu,\nu^{\prime}}(\mathbf{k})$ are the DFT and DMFT (defined
below) occupation matrices respectively, $\varepsilon_{\mathbf{k},\nu}$
is the eigen spectrum of the KS Hamiltonian with the density, corrected
by DMFT (in one-shot DFT+DMFT: $\varepsilon_{\mathbf{k},\nu}=\varepsilon_{\mathbf{k},\nu}^{DFT}$).
$E^{DC}$ is the double counting energy (defined in different approximations
in Appendix\ref{sec:app_Double}), while $E_{U}$ is the DMFT correlation
energy, which can be either calculated directly from the solver, as
the average of the interaction term, or via Galitskii-Migdal formula\citep{Migdal}:
\begin{equation}
E_{U}=\frac{1}{2}\sum_{\mathbf{R}}\mathrm{Tr}\left[G^{\mathbf{R}}(i\omega_{n})\Sigma^{\mathbf{R}}(i\omega_{n})\right].\label{eq:Migdal}
\end{equation}
By using a separation into a low-frequency numeric part and an analytic
sum of high-frequency tails, this summation can be accomplished efficiently.
We use a slightly modified version of the summation as explained in
Appendix\ref{sec:Matsubara_tails}.

Up to this point we did not specify the form of the localised basis
$\left|\chi_{m\mathbf{R}}\right\rangle $ and the formalism remained
general. In CASTEP, we use an already implemented LCAO basis, with
the radial part derived from pseudopotential\citep{DFTU_CASTEP},
which can be either norm-conserving or ultra-soft. In the case of
norm-conserving pseudopotentials, the states $\left|\chi_{m\mathbf{R}}\right\rangle $
are orthogonal by construction, while in the case of ultra-soft ones\citep{martin_2004}
these states are overlapping with an overlap matrix $S$:
\[
\left\langle \chi_{m^{\prime}\mathbf{R^{\prime}}}\left|\hat{S}\right|\chi_{m\mathbf{R}}\right\rangle =\delta_{m^{\prime},m}.
\]
This implies that the KS equation transforms from a standard eigenvalue
problem into a generalised one:
\[
\hat{H}_{k}^{KS}\left|\Psi_{\mathbf{k},\nu}\right\rangle =E_{\mathbf{k},\nu}\hat{S}\left|\Psi_{\mathbf{k},\nu}\right\rangle ,
\]
where we have introduced the KS eigenstates $\left|\Psi_{\mathbf{k},\nu}\right\rangle .$
The two cases (norm-conserving and ultra-soft pseudo-potentials) can
be unified by defining an overlap matrix in the norm-conserving case
to be identity matrix. In what follows, we will present the general
formalism, valid for both norm-conserving and ultra-soft pseudopotentials
used in CASTEP. It will become clear from what follows that the whole
formalism does not depend on $S$, provided that all the scalar products
are defined using $S$ as a metric. Next, we define the projectors
$P_{L,\nu}(\mathbf{k})$:
\begin{equation}
P_{L,\nu}(\mathbf{k})=\left\langle \chi_{L}\left|S\right|\Psi_{\mathbf{k},\nu}\right\rangle .\label{eq:Proj_NC}
\end{equation}
$P_{L,\nu}(\mathbf{k})$ are $S$-orthonormal to a high degree (in
both systems considered here the spilling factor was of the order
of $0.1\%$). In order to ensure the full $S$-orthogonality, we apply
L\"{o}wdin orthogonalisation procedure in the $S$-metric space.
From now on, we have two bases, spanning two different spaces: i)
Bloch space (indexed by $\mathbf{k},\nu$) and ii) localised basis
or ``correlated'' subspace (indexed by $L$). The two spaces are
connected by the projection procedure, also called up-folding (to
go from $\chi_{L}$ to $\Psi_{\mathbf{k},\nu}$):
\begin{equation}
\left|a_{\mathbf{k},\nu}\right\rangle =\sum_{L}P_{\nu,L}^{\star}(\mathbf{k})\left|b_{L}\right\rangle 
\end{equation}
 or down-folding (vice-versa):
\begin{equation}
\left|b_{L}\right\rangle =\sum_{\mathbf{k},\nu}P_{L,\nu}(\mathbf{k})\left|a_{\mathbf{k},\nu}\right\rangle .
\end{equation}
Here $\left|a_{\mathbf{k},\nu}\right\rangle $ is a vector living
in the Bloch space and $\left|b_{L}\right\rangle $ is a vector defined
in the space of ``correlated'' orbitals. For the current implementation
it is only important to have localised basis states on the ``correlated''
orbitals. The matrix $P_{L,\nu}(\mathbf{k})$ is, in general, a complex
rectangular matrix, satisfying the following condition:
\begin{align}
\sum_{k,\nu}P_{L,\nu}(\mathbf{k})P_{\nu,L^{\prime}}^{\star}(\mathbf{k}) & =\delta_{L,L^{\prime}}.
\end{align}
This condition is a consequence of completeness and $S$-orthogonality
of the KS eigen-basis, and the $S$-orthogonality (after L\"{o}wdin
orthogonalisation) of the ``correlated'' orbitals. Because both
Bloch and ``correlated'' spaces have the same metric, up- and down-folding
are accomplished ``as if there were no metric at all''. An important
consequence of this property stays in the fact that an up-folding
followed by a down-folding is an identity operation (in the ``correlated''
space), which guarantees that during DMFT iterations the charge is
conserved.

In the Bloch space the Bloch (or lattice) Green function can be obtained
from \eqref{eq:GF} by taking average over KS states $\left|\Psi_{\mathbf{k},\nu}\right\rangle $.
On the other hand, $G^{B}$ is a Fourier transform of $\langle\mathbf{r}\left|\hat{G}\right|\mathbf{r}\rangle$
into reciprocal space. In reciprocal space it takes the following
form:
\begin{align}
G_{\nu,\nu^{\prime}}^{B}(\mathbf{k},i\omega_{n}) & =\left(\left(i\omega_{n}+\mu-\varepsilon_{\mathbf{\mathbf{\mathbf{\mathbf{\kappa}}}},\nu}\right)\delta_{\nu,\nu^{\prime}}-\Sigma_{\nu,\nu^{\prime}}^{B}(\mathbf{k},i\omega_{n})\right)^{-1}\nonumber \\
 & =\mathrm{F.T.}\left[\langle\mathbf{R}\left|\hat{G}\right|\mathbf{0}\rangle\right].\label{eq:SigmaB}
\end{align}
Let us consider a correlated atom at position $\mathbf{R}$. The basis
functions in its ``correlated space'' are enumerated by index $m$.
As prescribed by the DMFT methodology, the local Green function at
that site is obtained from the Bloch one by down-folding and summation
over Brillouin zone:
\begin{equation}
G_{m,m^{\prime}}^{loc}(i\omega_{n})=\frac{1}{N_{\mathbf{k}}}\sum_{\nu,\nu^{\prime},\mathbf{k}}P_{m,\nu}(\mathbf{k})G_{\nu,\nu^{\prime}}^{B}(\mathbf{k},i\omega_{n})P_{\nu^{\prime},m^{\prime}}^{\star}(\mathbf{k}).
\end{equation}
 On the other hand, within the on-site Anderson impurity problem,
Dyson equation relates $G^{imp}$, $\Sigma^{imp}$ and the Weiss field
$\mathscr{G}_{0}$: 
\begin{equation}
\left[\mathscr{G}_{0}(i\omega_{n})\right]_{m,m^{\prime}}^{-1}=\Sigma_{m,m^{\prime}}^{imp}(i\omega_{n})+\left[G^{imp}(i\omega_{n})\right]_{m,m^{\prime}}^{-1}.
\end{equation}
The above equation serves as a definition for $\mathscr{G}_{0}$ by
making the fundamental DMFT assumption: $G^{imp}=G^{loc}$ (and $\Sigma^{imp}=\Sigma^{loc}$).
$\mathscr{G}_{0}$ will be used by the impurity solver in the next
step. Alternatively, one can use the hybridisation $\Delta(i\omega_{n})$
instead of $\mathscr{G}_{0}$:
\begin{align}
\Delta_{m,m^{\prime}}(i\omega_{n}) & =i\omega_{n}-\epsilon_{m,m^{\prime}}+\mu-\left[\mathscr{G}_{0}(i\omega_{n})\right]_{m,m^{\prime}}^{-1}
\end{align}
Here $\epsilon_{m,m^{\prime}}$ is the local impurity energy matrix,
obtained by down-folding the KS Hamiltonian onto ``correlated space''
of the given correlated atom:
\begin{equation}
\epsilon_{m,m^{\prime}}=\frac{1}{N_{\mathbf{k}}}\sum_{\mathbf{k},\nu}P_{m,\nu}(\mathbf{k})\varepsilon{}_{k,\nu}^{KS}P_{\nu,m^{\prime}}^{\star}(\mathbf{k}).
\end{equation}

The outcome of the impurity solver is the new impurity self-energy
denoted as $\Sigma_{m,m^{\prime}}^{imp}(i\omega_{n})$. It is subsequently
up-folded into the Bloch subspace (after the subtraction of the double-counting
corrections $V_{m,m^{\prime}}^{DC}$):
\begin{equation}
\Sigma_{\nu,\nu^{\prime}}^{B}(\mathbf{k},i\omega_{n})=P_{\nu,m}^{\star}(\mathbf{k})\left(\Sigma_{m,m^{\prime}}^{imp}(i\omega_{n})-V_{m,m^{\prime}}^{DC}\right)P_{m^{\prime},\nu^{\prime}}(\mathbf{k}).
\end{equation}
Thus up-folded Bloch self-energy acquires $k$-dependence. $\Sigma^{B}$
is then inserted into \eqref{eq:SigmaB} and the calculations proceed
until the convergence on chemical potential and self-energy is reached
with a given tolerance.

At convergence, the system's properties can be evaluated: total energy
from \eqref{eq:E_tot}, and, in principle, any single particle properties
from the Bloch Green function. For example, the DFT+DMFT occupation
matrix $N_{\nu,\nu^{\prime}}(k)$ (which is not diagonal, unlike in
conventional DFT) is obtained from $G_{\nu,\nu^{\prime}}^{B}(\mathbf{k},i\omega_{n})$
as:
\begin{align}
N_{\nu,\nu^{\prime}}(\mathbf{k}) & =T\sum_{n}G_{\nu,\nu^{\prime}}^{B}(\mathbf{k},i\omega_{n})e^{i\omega_{n}0^{+}},
\end{align}
and hence the total number of electrons in the unit cell, used to
fix the chemical potential $\mu$, is given by:
\begin{equation}
N_{e}=\frac{1}{N_{\mathbf{k}}}\sum_{\nu,\mathbf{k}}N_{\nu,\nu}(\mathbf{k}).
\end{equation}
The spectral density $A(\mathbf{k},\omega)$ (in real frequency) is
derived from analytically continued (see details in the next subsection)
$G^{B}$ as:
\begin{equation}
A_{\nu,\nu}(\mathbf{k},\omega)=-\frac{1}{\pi}\mathrm{Im}G_{\nu,\nu}^{B}(\mathbf{k},\omega),
\end{equation}
while the total DOS $D(\omega)$ is in turn obtained from $A_{\nu,\nu}(\mathbf{k},\omega)$
by integrating over Brillouin zone:
\begin{equation}
D(\omega)=\frac{1}{N_{k}}\sum_{k,\nu}A_{\nu,\nu}(\mathbf{k},\omega).
\end{equation}
One can also calculate the partial DOS derived from the impurity Green
function:
\begin{equation}
D_{imp}(\omega)=-\frac{1}{\pi}\sum_{m}\mathrm{Im}G_{m,m}^{imp}(\omega).
\end{equation}

\begin{figure}
\includegraphics[bb=0bp 0bp 1287bp 861bp,clip,width=1\columnwidth]{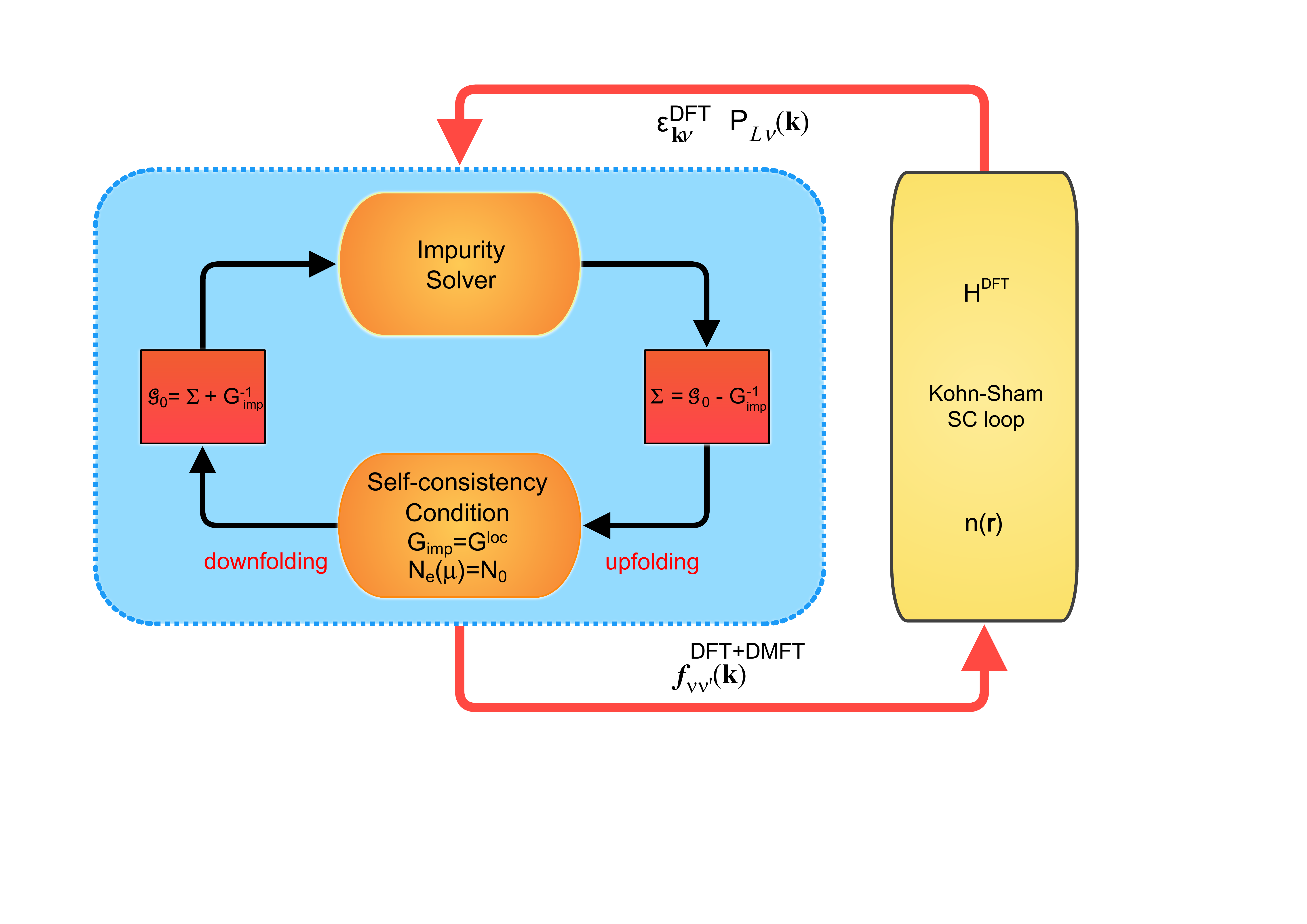}

\caption{\label{flowchart}(Color online) DFT+DMFT execution flowchart, containing both the
inner self-consistency loop (DMFT at fixed charge density) as well
as the outer one (Kohn-Sham equations at DFT+DMFT charge density).}

\end{figure}

To conclude this subsection, we summarise the program work-flow. The
execution proceeds as follows:

\renewcommand{\labelenumi}{\roman{enumi})}
\begin{enumerate}
\item The electronic density is converged at the DFT level
\item An initial guess for self-energy $\Sigma_{m,m^{\prime}}^{\mathbf{R}}$
is made, which is subsequently up-folded into Bloch space: $\Sigma^{B}$
\item Chemical potential $\mu$ is determined at fixed $\Sigma^{B}$
\item $\mathscr{G}_{0}(i\omega_{n})$ (or $\Delta(i\omega_{n})$) is formed
by down-folding $G^{B}$
\item Call of impurity solver updates $\Sigma_{m,m^{\prime}}^{\mathbf{R}}$
\item Up-folding $\Sigma_{m,m^{\prime}}^{\mathbf{R}}$ gives new $\Sigma^{B}$
\item If the convergence for $\mu$ and $\Sigma^{B}$ is not reached, go
to iii)
\item If full charge self-consistency is requested, update the charge density
$n(\mathbf{r})$ and go to i)
\item Compute system's properties within DMFT.
\end{enumerate}
This work-flow is illustrated in Fig.\ref{flowchart}.

\subsection{Solvers}

It is evident from the previous subsection that the central point
of DMFT method is the solution of the impurity problem. This is normally
accomplished by the so called impurity solver. Several methods have
been developed in the past. At present, we implement in CASTEP three
impurity solvers:
\begin{enumerate}
\item Hubbard I (see e.g. Ref.\onlinecite{HubbardI,Pourovskii_2007})
\item Continuous Time Quantum Monte Carlo with Hybridization expansion CT-HYB
available through TRIQS package\citep{TRIQS_REF}
\item Exact Diagonalisation with Cluster Perturbation Theory (ED-CPT) solver\citep{Weber2012}. 
\end{enumerate}
Each of these solvers has its advantages and deficiencies which we
list shortly below. Within Hubbard I approximation the impurity is
treated as an isolated atom (atomic limit) and the hybridization with
the bath is totally neglected. The Weiss field in Hubbard I can be
expressed as: $\mathscr{G}_{0}^{-1}=i\omega_{n}+\mu-\epsilon$. Of
course, such an approximation is very crude, but might be acceptable
for strongly localised orbitals (e.g. $f$-shells in rare-earth elements).
Moreover, an important advantage of Hubbard I consists in its ability
to work on both real and imaginary frequency axes, allowing analytic
continuation to be avoided. Finally, it is fast and free from statistical
bias, which allows to use it for quick tests and for total energy
and forces calculations.

In contrast to the Hubbard I method, in the case of density-density only interactions,
a CT-HYB solver offers a numerically exact solution to
the impurity problem with a given Weiss field $\mathscr{G}_{0}(i\omega_{n})$
at a reasonable computational cost. As is evident from its name, CT-HYB
builds its perturbation expansion in powers of hybridisation and therefore
could require more resources in case of a strongly hybridised impurity.
The output of CT-HYB solver is the self-energy in imaginary frequency,
which means that some routine for analytic continuation is needed
to obtain the real-axis results. In CASTEP, we use the Pade approximation\citep{Pade_REF}
with the calculations using arbitrary precision arithmetic\citep{FMlib}
in order to face the problem of precision loss inherent to the Pade
approximation.

Finally, the ED-CPT solver is a kind of a compromise between the strengths
and weaknesses of the Hubbard I and CT-HYB solvers. Like the CT-HYB
solver, it avoids truncating the Weiss field. Like the Hubbard I solver,
it can work on either the real or the imaginary axis, it does not
introduce any stochastic error, and it works well in strongly hybridized
problems. The ED-CPT solver does suffer a systematic error caused
by bath discretisation, when the Weiss field, having the meaning of
an infinite bath Green function, is approximated by a model function
with a finite number of bath sites. However this problem is mitigated
by the use of cluster perturbation theory, and is further decreased
when using modern HPC computational resources (including GPU cards)
which allows the treatment of systems with up to $18$ single-orbital
sites; this is quite close to the maximum number of sites tractable
with exact diagonalisation, due to the exponential growth of the Hilbert
space with the number of sites\citep{Koch2014}.

\begin{table}
\begin{tabular}{|c|c|c|}
\hline 
$\mathbf{\gamma-Ce}$ & $a$ (\angstrom) & $B_{0}$ (GPa)\tabularnewline
\hline 
\hline 
Experiment\citep{Amadon2012} & $5.17$ & $19/21$\tabularnewline
\hline 
Present work LDA+DMFT & $4.95$ & $30$\tabularnewline
\hline 
PAW/LDA+DMFT\citep{Amadon2012} & $4.98$ & $38$\tabularnewline
\hline 
ASA/LDA+DMFT\citep{Amadon2012} & $4.91$ & $50$\tabularnewline
\hline 
\multicolumn{1}{c}{} & \multicolumn{1}{c}{} & \multicolumn{1}{c}{}\tabularnewline
\hline 
$\mathbf{Ce_{2}O_{3}}$ & $a$ (\angstrom) & $B_{0}$ (GPa)\tabularnewline
\hline 
\hline 
Experiment\citep{Amadon2012} & $3.89$ & $111$\tabularnewline
\hline 
Present work LDA+DMFT & $3.81$ & $164$\tabularnewline
\hline 
PAW/LDA+DMFT\citep{Amadon2012} & $3.76$ & $170$\tabularnewline
\hline 
ASA/LDA+DMFT\citep{Amadon2012} & $3.79$ & $160$\tabularnewline
\hline 
\multicolumn{1}{c}{} & \multicolumn{1}{c}{} & \multicolumn{1}{c}{}\tabularnewline
\hline 
$\mathbf{SmTe}$ & $a$ (\angstrom) & $B_{0}$ (GPa)\tabularnewline
\hline 
\hline 
Experiment\citep{BENEDICT_1993} & $6.58$ & $43.5$\tabularnewline
\hline 
Present work LDA+DMFT & $6.30$ & $54.2$\tabularnewline
\hline 
Present work LDA & $6.09$ & $65.5$\tabularnewline
\hline
\end{tabular}

\caption{\label{tab:Comparison}Comparison of the lattice constant $a$ and
bulk modulus $B_{0}$ of $\gamma$-Ce, Ce$_{2}$O$_{3}$ and SmTe calculated
within CASTEP's DFT+DMFT implementation with experimental data as
well as with theoretical results of Ref.\onlinecite{Amadon2012}.}
\end{table}

\section{\label{sec:Examples}Examples}

\subsection{Structural properties of $\gamma-$Ce}

Elemental cerium is well known for having several phases ($\alpha$,
$\beta$, $\gamma$, $\delta$, $\alpha^{\prime}$, $\alpha^{\prime\prime}$
etc.), for a review, see Ref.\onlinecite{ReviewCe}. The most puzzling
and the most studied phase transition is the $\alpha-\gamma$ iso-structural
transition, which is accompanied by a $15\%$ volume collapse at room
temperature. It is believed that the lattice structure in both $\alpha$
and $\gamma$ phases is the same (fcc), the lattice constant being
the only difference. Within the Mott localisation theory of $\alpha-\gamma$
transition in Ce, the transition is viewed as a localisation of $f$
electrons in $\gamma$ phase, while in $\alpha$ phase they remain
itinerant\citep{Casula_2015}. We focus here on $\gamma$ phase. Its
lattice constant is underestimated within LDA by $13\%$ (see below),
which is due to the inability of the LDA to adequately describe the
localisation effects. Post-DFT methods such as DFT$+$U and DFT$+$DMFT
improve the agreement with the experiment, although could not recover
$100\%$ of the experimental value\citep{Amadon2012}.

\begin{figure}
\includegraphics[bb=0bp 0bp 407bp 599bp,angle=270,width=1\columnwidth]{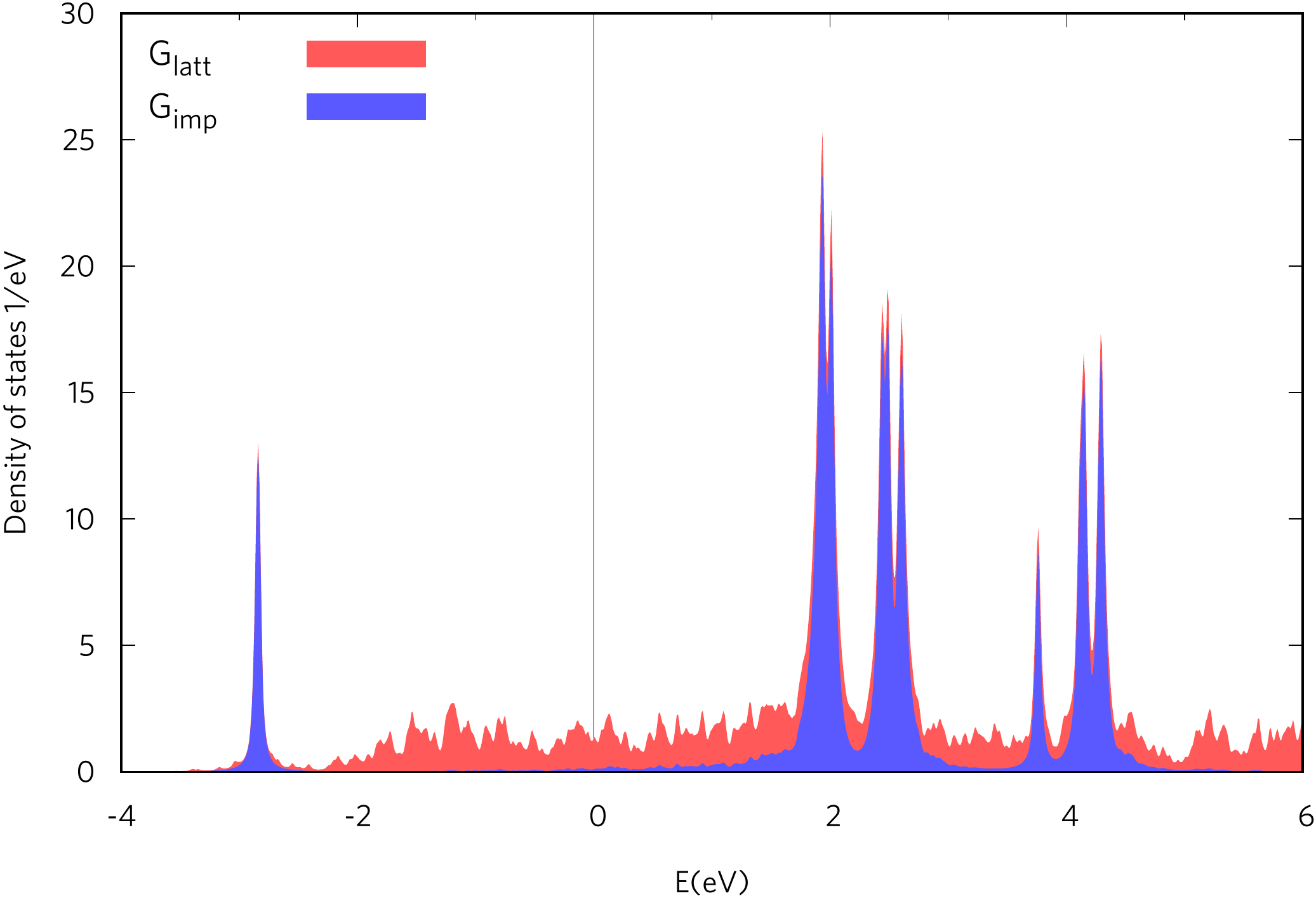}

\caption{\label{fig:gamma-Ce-Full-DOS-1}(Color online). Density of states
of $\gamma$-Cerium, calculated by CASTEP's DFT+DMFT implementation
and using the Hubbard I impurity solver. $G_{imp}$ labels the impurity
Green function derived DOS of Ce $f$-states, and $G_{latt}$ labels
the $G^{B}(k,\omega)$ derived DOS.}
\end{figure}

We have used here a $15\times15\times15$ Monkhorst-Pack $k$-point
mesh\citep{MPgrid} (equivalent to $k$-point spacing of $0.02$ \angstrom$^{-1}$),
and the rhombohedral unit cell with $a_{exp}=5.161\angstrom$ (experimental
value), having a primitive unit cell volume of $34.37\angstrom{}^{3}$.
For Ce, we have used CASTEP's internally generated scalar relativistic ultra-soft pseudopotential
(C9 set) and the following values of Hubbard $U$ and $J$: $U=6$eV and $J=0.7$eV. The simulations
were carried out at $T=0.02$eV.
The plane-wave
basis cut-off was automatically determined to be $359$eV. In Fig.\ref{fig:gamma-Ce-Full-DOS-1},
we report the density of states calculated at the experimental lattice
constant $a_{exp}$ using the Hubbard I solver.

It can be clearly
seen that the CASTEP+DMFT implementation captures the overall shape
of the Density of States (DOS) very well as compared to Fig.5a of
Ref.\onlinecite{Pourovskii_2007} and to Ref.\onlinecite{TRIQS_tut},
while our results appear to be shifted by approximately $0.5$eV,
which can be ascribed to the difference in treatment of projections:
namely, we have used the whole energy range of KS eigenstates, as
opposed to Refs.\onlinecite{Pourovskii_2007,TRIQS_tut}, where an
energy window was imposed. The imposition of an energy window implies
neglecting the change of the electronic density from the energy regions
beyond the window, which may lead to shifts of the chemical potential.
In $\gamma$-Ce, the application of DMFT leads to the opening of a
gap in the $f$ states, being the residual spectral weight due to
other orbital moments ($d$- and $p$-states). It is these residual
states in the Bloch Green function, strongly dependent on the projection
procedure, which eventually determine whether the chemical potential
of the insulating system stays at the top of valence band or at the
bottom of conduction one.
Finally, in our calculations there appear extra high energy peaks around $4$eV due to
Ce $f$-states as compared to Ref.~\onlinecite{Pourovskii_2007,Amadon2012}. We have
checked that the origin of these peaks is due to a finite Hund's coupling $J$ used
in our calculations, as opposed to Ref.~\onlinecite{Pourovskii_2007,Amadon2012}, where $J=0$ was used.

\begin{figure}
\includegraphics[bb=0bp 0bp 419bp 622bp,angle=270,width=1\columnwidth]{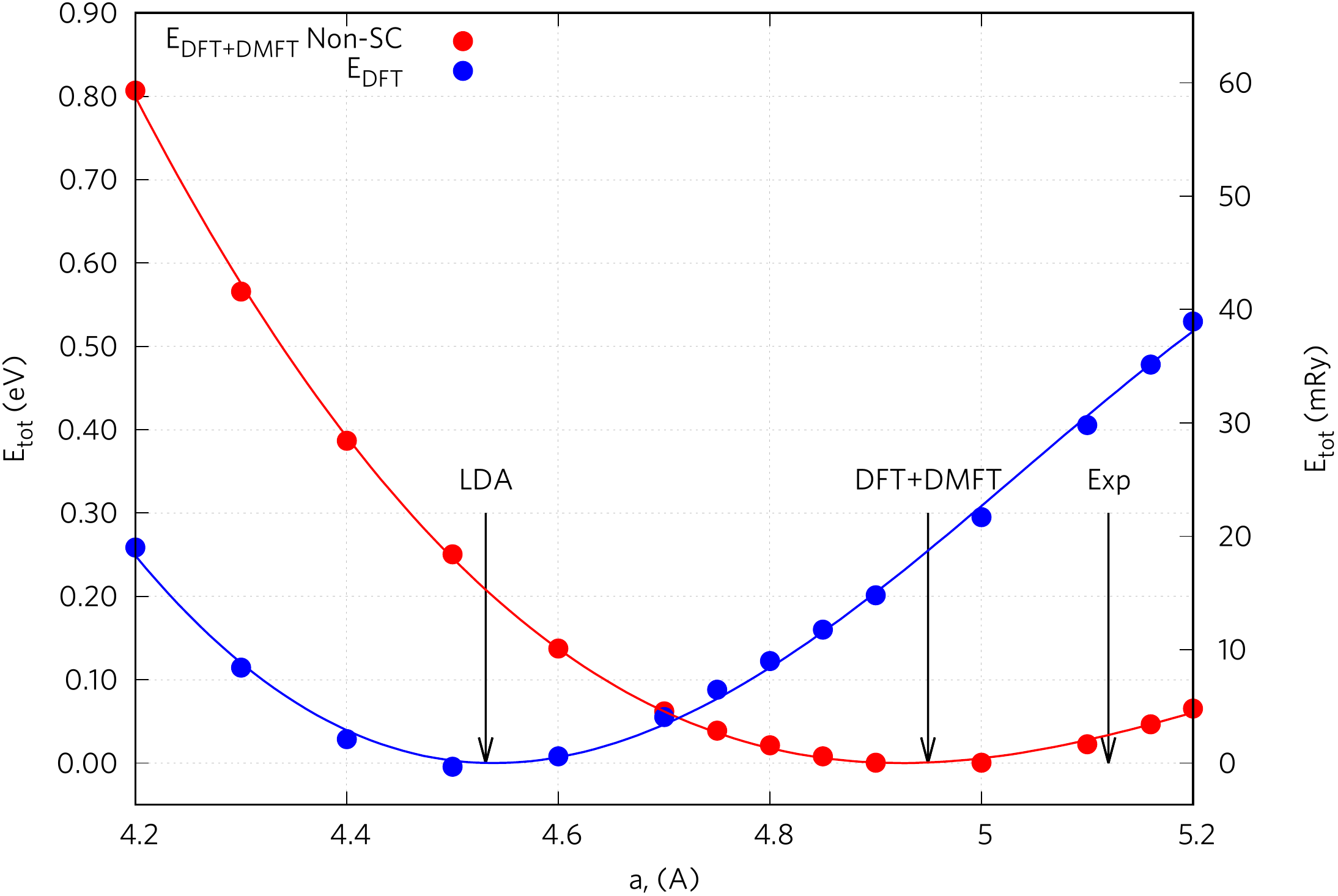}

\caption{\label{fig:Ce_E_vs_V}(Color online). $\gamma$-Cerium's total energy
$E_{tot}$ as a function of lattice constant $a$, calculated both
with DFT and with DFT+DMFT. Arrows show the experimental, DFT, and
DFT+DMFT values of the equilibrium lattice constant. Curves show Birch-Murnaghan
fits to the calculated points.}
\end{figure}

We have also studied the total energy as a function of volume, shown
in Fig.\ref{fig:Ce_E_vs_V} and Tab.\ref{tab:Comparison}. One can
notice a very good qualitative and quantitative agreement of our results
with those of Ref.\onlinecite{Pourovskii_2007}: while the DFT energy
minimum is realised at $a=4.50$\angstrom (not shown), taking into
account the localisation effects within DFT+DMFT, shifts the minimum
to $a=4.95$\angstrom, a result slightly closer to the experimental
value than that of Ref.\onlinecite{Pourovskii_2007}. It is interesting
to note that among five contributions to the total energy expression,
only two are active in the case of Ce, namely the second and the third
terms in Eq.\eqref{eq:E_tot}. Indeed, it is argued in Ref.\onlinecite{Pourovskii_2007}
that for the Hubbard I solver applied to Ce $f$-shell, an integer
occupation with one electron should be used independently of the lattice
constant, and in these circumstances $E_{U}=0$, while $E^{DC}$ does
not depend on the lattice constant. We remind that everywhere throughout
this paper we performed DFT+DMFT calculations with fixed charge. We
have applied the Fully Localised Limit (FLL) type of double counting
corrections (see Appendix\ref{sec:app_Double}).

Another structural property which is known to be corrected within
DFT+DMFT is the bulk modulus $B_{0}$. By fitting the Birch-Murnaghan\citep{Murnaghan44,Birch47,Hebbache_2004}
equations of state to the energy versus volume curves of Fig.\ref{fig:Ce_E_vs_V}
we obtain an estimate for $B_{0}$ which is in line with the predictions
of Ref.\onlinecite{Amadon2012}, as shown in Table \ref{tab:Comparison}.
Moreover, even though in general DFT+DMFT systematically overestimates
$B_{0}$, we can see from Table \ref{tab:Comparison} that our results
are closer to the experimental ones (less overestimating). This is
probably because of the difference in the underlying DFT method, as
can be seen in Table \ref{tab:Comparison}, where the results from Ref.\onlinecite{Amadon2012}
for PAW/LDA+DMFT and ASA/LDA+DMFT are clearly different, although the DMFT
treatment was identical.

\subsection{Structural properties of cerium sesquioxide}

\begin{figure}
\includegraphics[bb=0bp 0bp 419bp 617bp,angle=270,width=1\columnwidth]{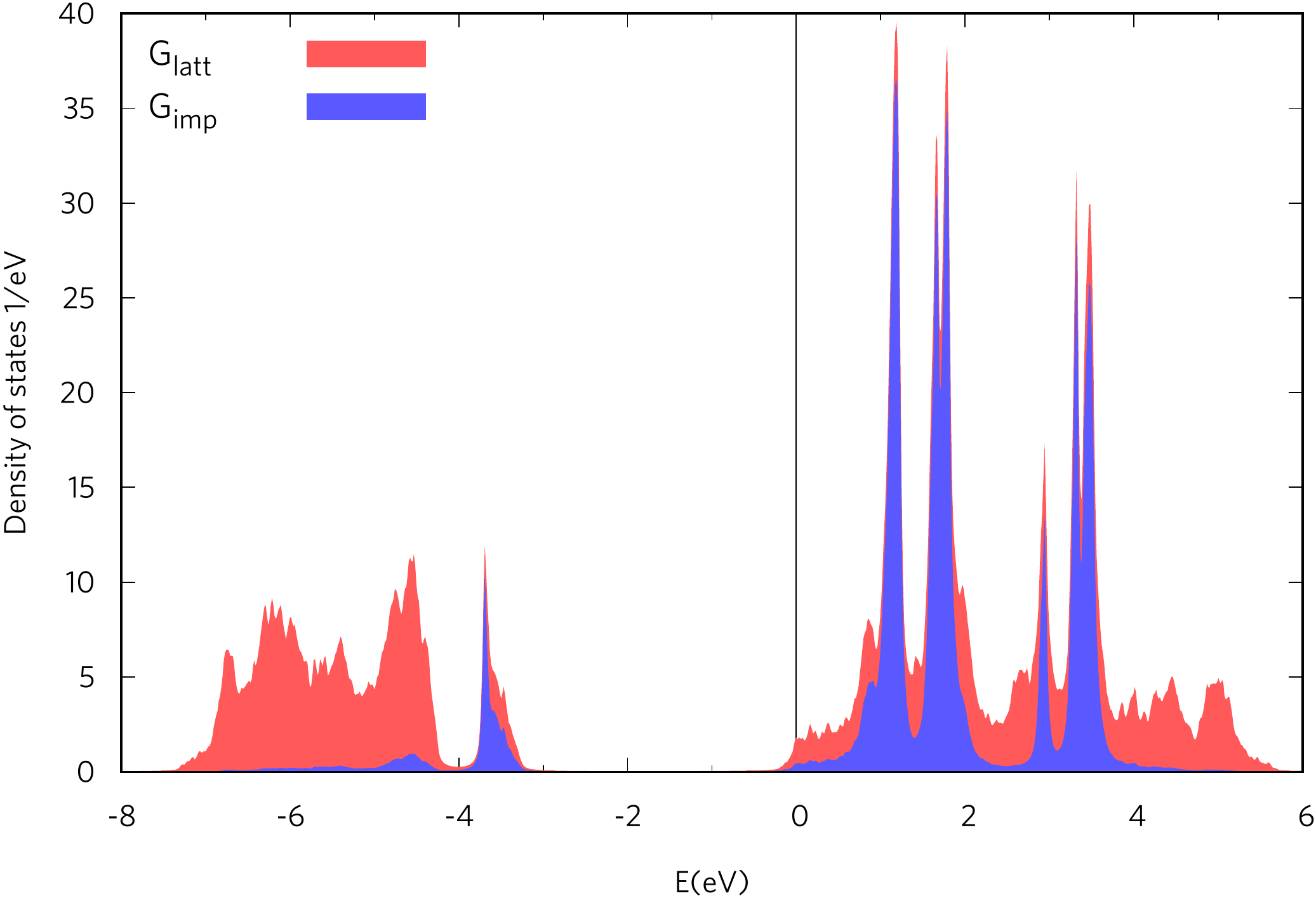}

\caption{\label{fig:DOS_Ce2O3}(Color online). Density of states of Ce$_{2}$O$_{3}$
calculated by CASTEP's DFT+DMFT implementation and using the with
Hubbard I solver. $G_{imp}$ labels the impurity Green function derived
DOS of Ce $f$-states, and $G_{latt}$ labels the $G^{B}$(k,$\omega$)
derived DOS.}
\end{figure}

Cerium sesquioxide Ce$_{2}$O$_{3}$ has been studied for a long time\citep{Andersson_2007,Fabris_2005,Singh_2006,Loschen_2007}.
It is known to be an anti-ferromagnetic insulator with N\'{e}el temperature
of $T_{N}=9K$ and a gap of $2.4$eV. DFT+DMFT calculations in the
literature normally address the high-temperature paramagnetic phase,
so to benchmark our implementation we also set the temperature to
$T=0.02$eV. Ce$_{2}$O$_{3}$ crystallises in a hexagonal unit cell
with space group $P\bar{3}m1$. The experimental parameters for the
unit cell are: $a=3.89$$\angstrom$ and $c/a=1.557$, with the Wyckoff
positions\citep{Wyckoff67}: Ce $2d$ $\left(\frac{1}{3},\frac{2}{3},0.24543\right)$,
O $2d$ $(\frac{1}{3},\frac{2}{3},0.6471)$, O $1a$ $(0,0,0)$. We
have used the same Ce pseudopotential as in the previous subsection,
and CASTEP's on-the-fly generated ultra-soft pseudopotential for oxygen (C9 set), and a $17\times17\times9$
Monkhorst-Pack $k$-point mesh (equivalent to $k$-point spacing of
approximately $0.02$ \angstrom$^{-1}$). The plane-wave basis cut-off
was automatically determined to be $653$eV. The results for Ce$_{2}$O$_{3}$
density of states at the experimental geometry are shown in Fig.\ref{fig:DOS_Ce2O3}
and exhibit excellent agreement with the reference calculations of
Ref.\onlinecite{Pourovskii_2007}. As before, the DMFT calculations
were performed with: a Hubbard I solver; and a fixed occupancy of
$n=1$ per Ce atom (in the sense explained in Ref.\onlinecite{Pourovskii_2007})
within the FLL double-counting scheme. The result of the application
of charge non-self-consistent DMFT in Ce$_{2}$O$_{3}$ is the opening
of a $3$eV gap in the total density of states (while taking into
account the charge self-consistency manages to reproduce the experimental
gap of $2.4$eV, according to the results of Ref.\onlinecite{Pourovskii_2007}).
The quantitative agreement of our results with those of Ref.\onlinecite{Pourovskii_2007}
is excellent, except for the shift of the chemical potential in the
gap, which can be attributed to the difference in the procedure of
fixing the total electronic density, as explained in the previous
subsection.

\begin{figure}
\includegraphics[bb=0bp 0bp 422bp 630bp,angle=270,width=1\columnwidth]{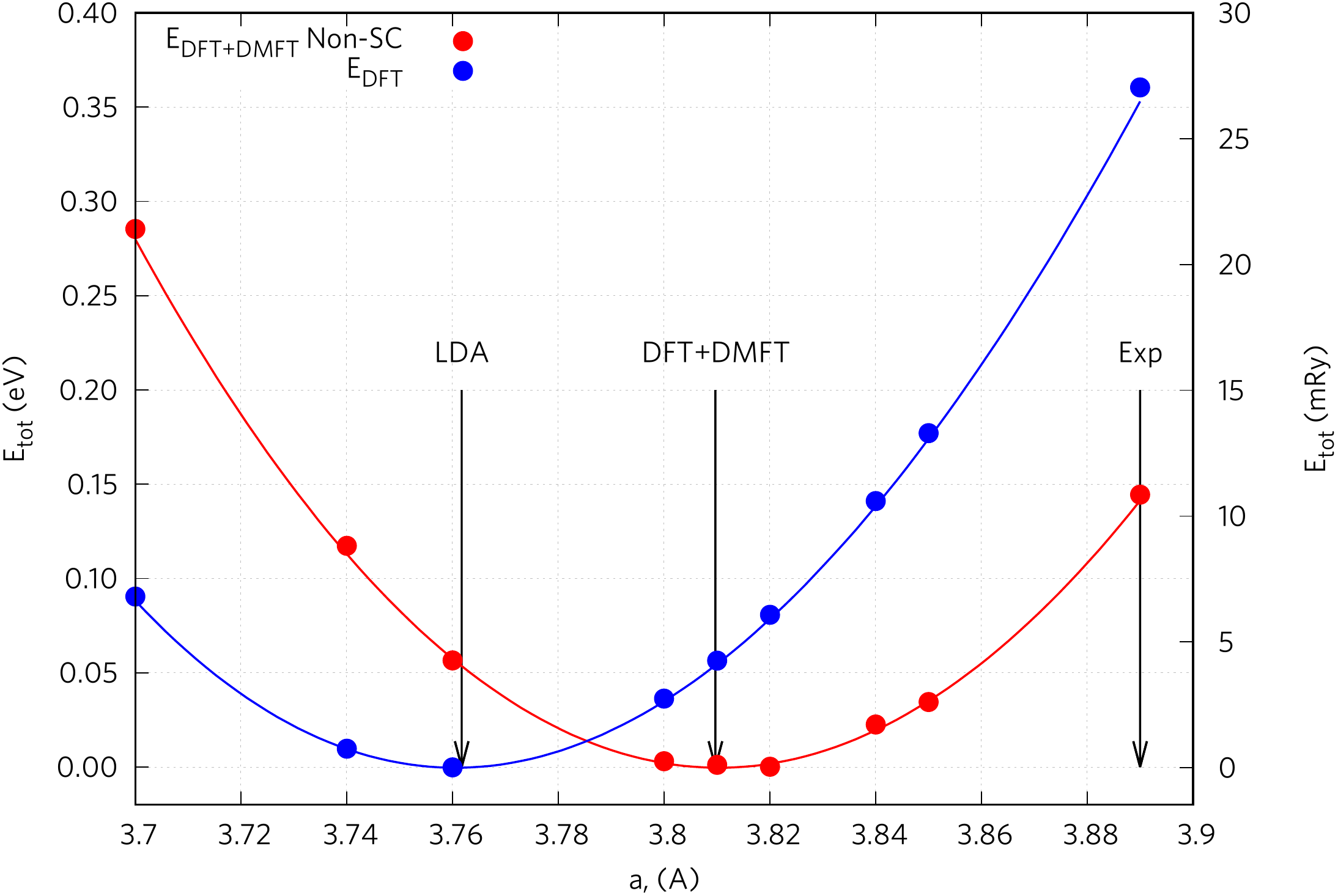}

\caption{\label{fig:Ce2O3_E_tot}(Color online). Ce$_{2}$O$_{3}$'s total
energy $E_{tot}$ as a function of lattice constant $a$, calculated
both with DFT and with DFT+DMFT. Arrows show the experimental, DFT,
and DFT+DMFT values of the equilibrium lattice constant. Curves show
Birch-Murnaghan fits to the calculated data points.}
\end{figure}

The same level of agreement with the reference calculations is exhibited
by our total energy calculations, as shown in Fig.\ref{fig:Ce2O3_E_tot}
and Table \ref{tab:Comparison}. In doing these calculations, we maintained
the ratio $c/a$ as well as the internal positions of the atoms in
the unit cell fixed, while changing $a$. Compared to DFT calculations,
which stabilise the unit cell around $a=3.76$\angstrom, the DMFT
energy minimum is at a larger value of $3.81$\angstrom, which is
very close to the results of Refs.\onlinecite{Pourovskii_2007,Amadon2012}.
Moreover, our result for the lattice constant $a$ is somewhat closer
to the experimental value, while our $B_{0}$ is between the two results
of Ref.\onlinecite{Amadon2012}.

\subsection{Structural properties of SmTe}


In order to demonstrate the capabilities of the implementation further, we apply it to a study of the equation
of state of samarium telluride (SmTe). To the best of
our knowledge this is the first study of SmTe in the literature with
DFT+DMFT. 
We have used here a $19\times19\times19$ Monkhorst-Pack $k$-point
mesh\citep{MPgrid} (equivalent to $k$-point spacing of $0.02$ \angstrom$^{-1}$),
and the face-centered cubic unit cell with two atoms (Sm at $(0,0,0)$ and Te at
$(\frac{1}{2},\frac{1}{2},\frac{1}{2})$). We have scanned the values of cubic lattice constants
from $a=5.5$\angstrom\; to $a=6.8$\angstrom.
For $f$-electrons on Sm, we have used $U=6.1$eV and $J=0.835$eV. For Sm and Te, we have used CASTEP's
internally generated scalar relativistic ultra-soft pseudopotentials (C9 set). The plane-wave basis
cut-off was $425$eV. 

\begin{figure}
\includegraphics[bb=0bp 0bp 419bp 617bp,angle=270,width=1\columnwidth]{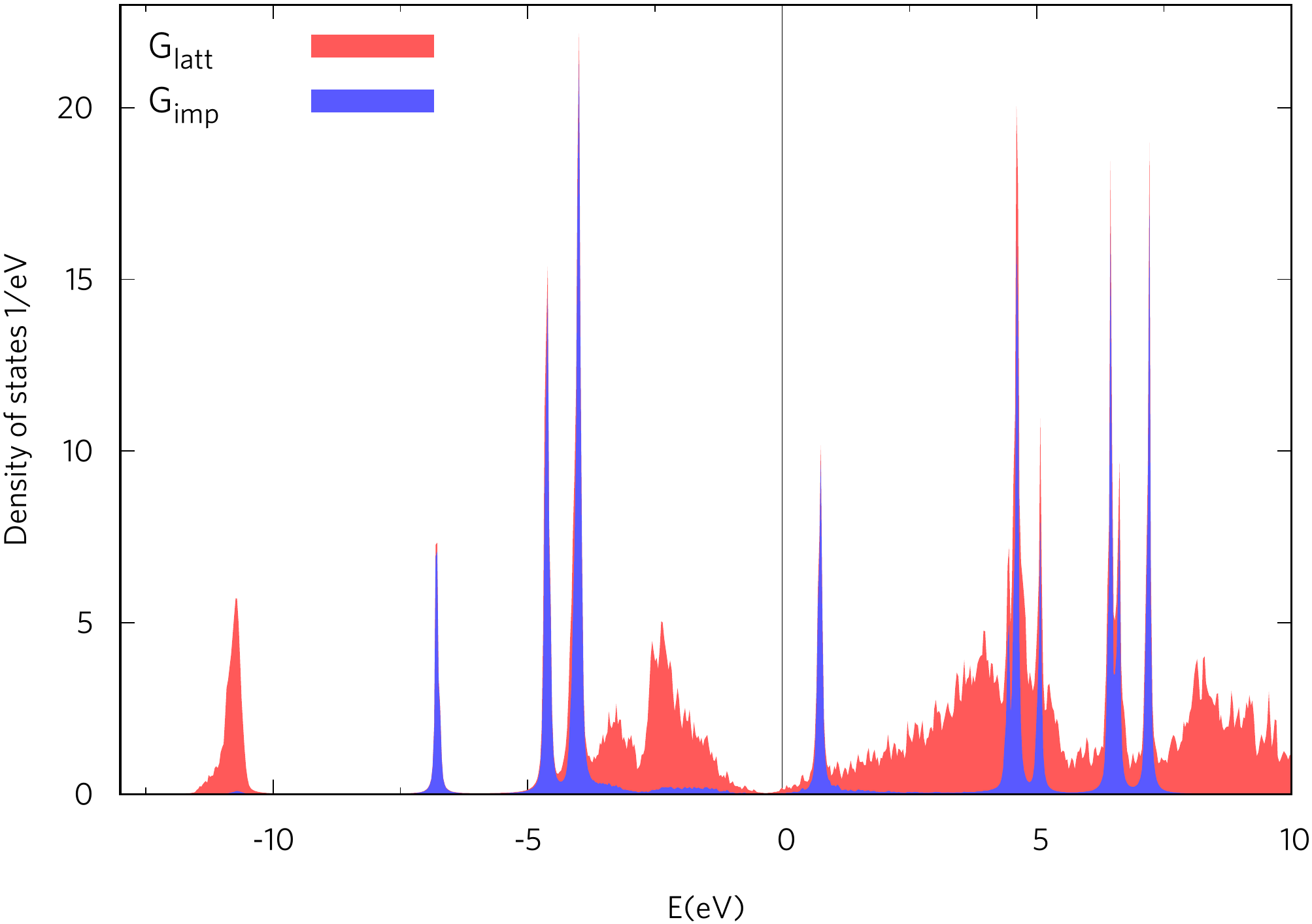}

\caption{\label{fig:DOS_SmTe}(Color online). Density of states of SmTe 
calculated by CASTEP's DFT+DMFT implementation and using the with
Hubbard I solver. $G_{imp}$ labels the impurity Green function derived
DOS of Sm $f$-states, and $G_{latt}$ labels the $G^{B}$(k,$\omega$)
derived DOS.}
\end{figure}

In Fig.\ref{fig:DOS_SmTe},
we report the density of states calculated at the value of $a$ corresponding to a minimum of
$E_{tot}$ within DFT+DMFT ($a_{DMFT}=6.3$\angstrom) using the Hubbard I solver. It can be seen that
the effect of improved treatment of the electronic correlations of $f$-electrons on Sm is to open a
gap in the $f$ states and to remove them from the Fermi level, so that the system becomes a
semiconductor in accordance with the experimental findings\cite{}.

\begin{figure}
\includegraphics[bb=0bp 0bp 422bp 630bp,angle=270,width=1\columnwidth]{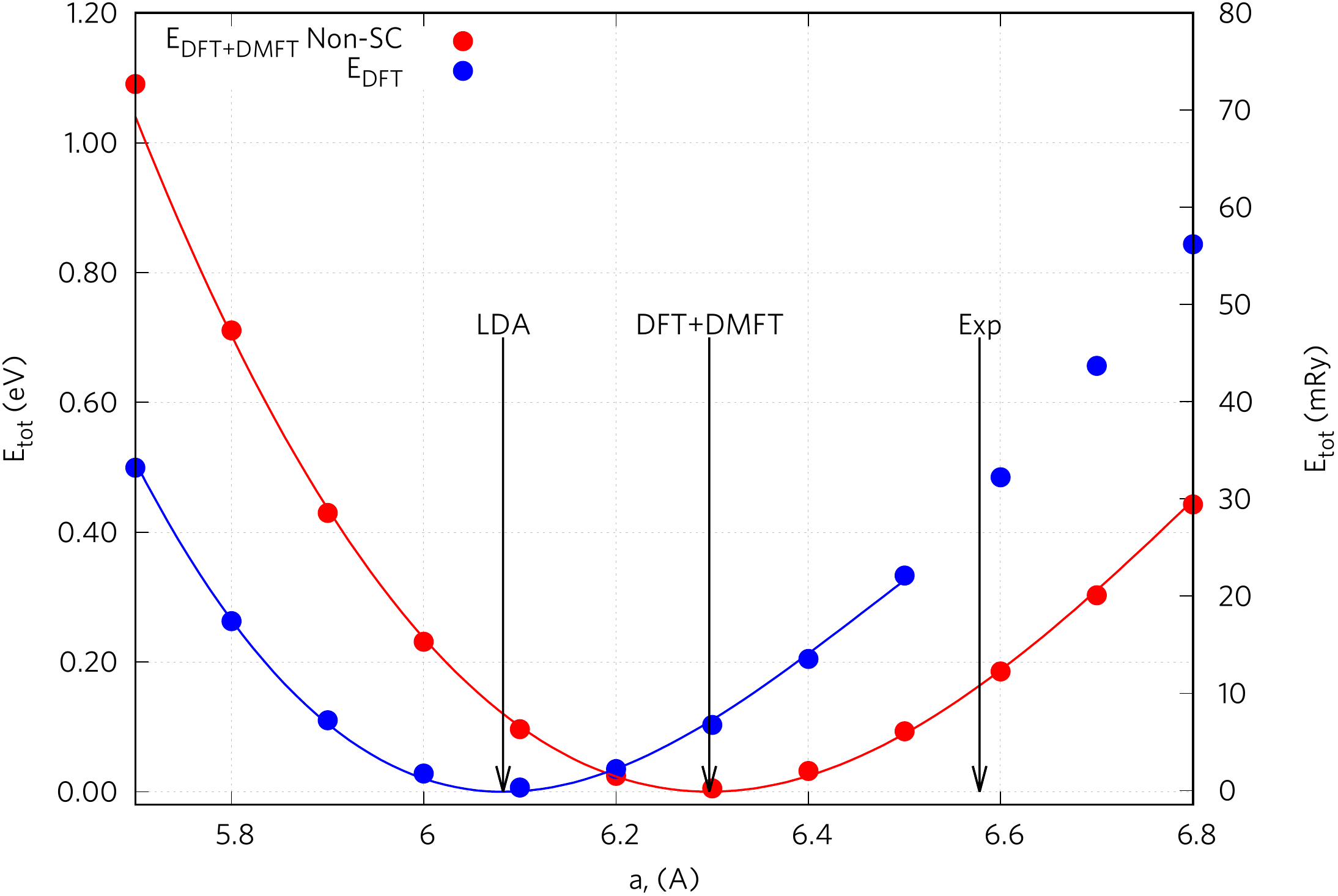}

\caption{\label{fig:SmTe_E_tot}(Color online). SmTe's total
energy $E_{tot}$ as a function of lattice constant $a$, calculated
both with DFT and with DFT+DMFT. Arrows show the experimental, DFT,
and DFT+DMFT values of the equilibrium lattice constant. Curves show
Birch-Murnaghan fits to the calculated data points.}
\end{figure}

Standard LDA underestimates the equilibrium lattice constant of SmTe due
to its inability to properly treat the
Sm $f$-orbitals' localisation, as can be seen from Fig.\ref{fig:SmTe_E_tot}. Inclusion
of the localisation effects within our DFT+DMFT implementation increases the equilibrium $a$. The
improvement with respect to LDA is as follows: LDA mismatch is $7\%$, while DFT+DMFT mismatch is
$4\%$. The same type of improvement is observed for bulk modulus as can
be seen from Table~\ref{tab:Comparison}: LDA overestimates $B_0$ by $51\%$,
while DFT+DMFT estimate is closer to the experimental value ($25\%$ of
error). It is evident that our implementation of DFT+DMFT significantly
improves the agreement of strongly correlated materials simulations with
the experiment.

\section{\label{sec:Forces}Calculation of forces in cerium sesquioxide}

\begin{figure}
\includegraphics[angle=270,width=1\columnwidth]{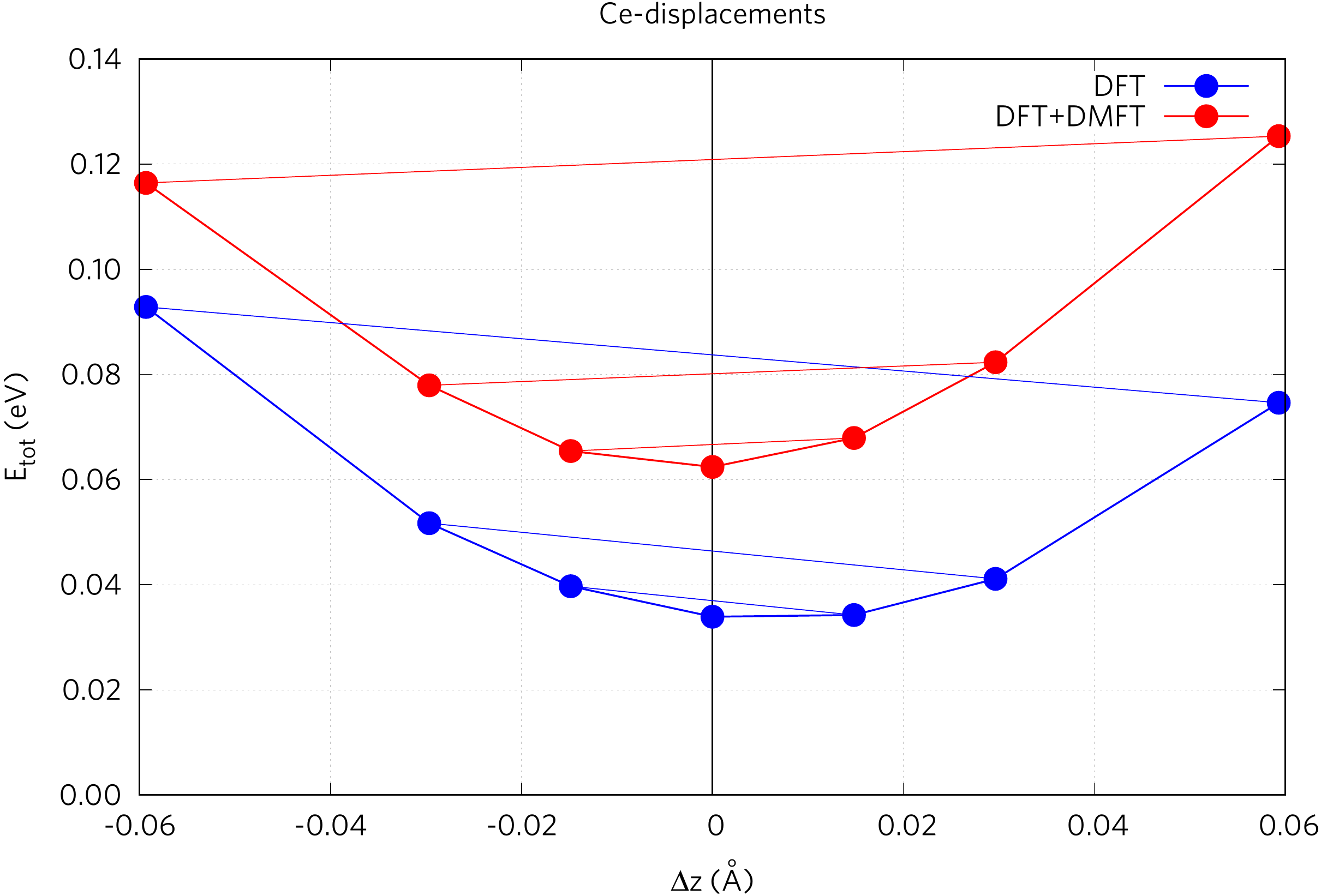}

\includegraphics[angle=270,width=1\columnwidth]{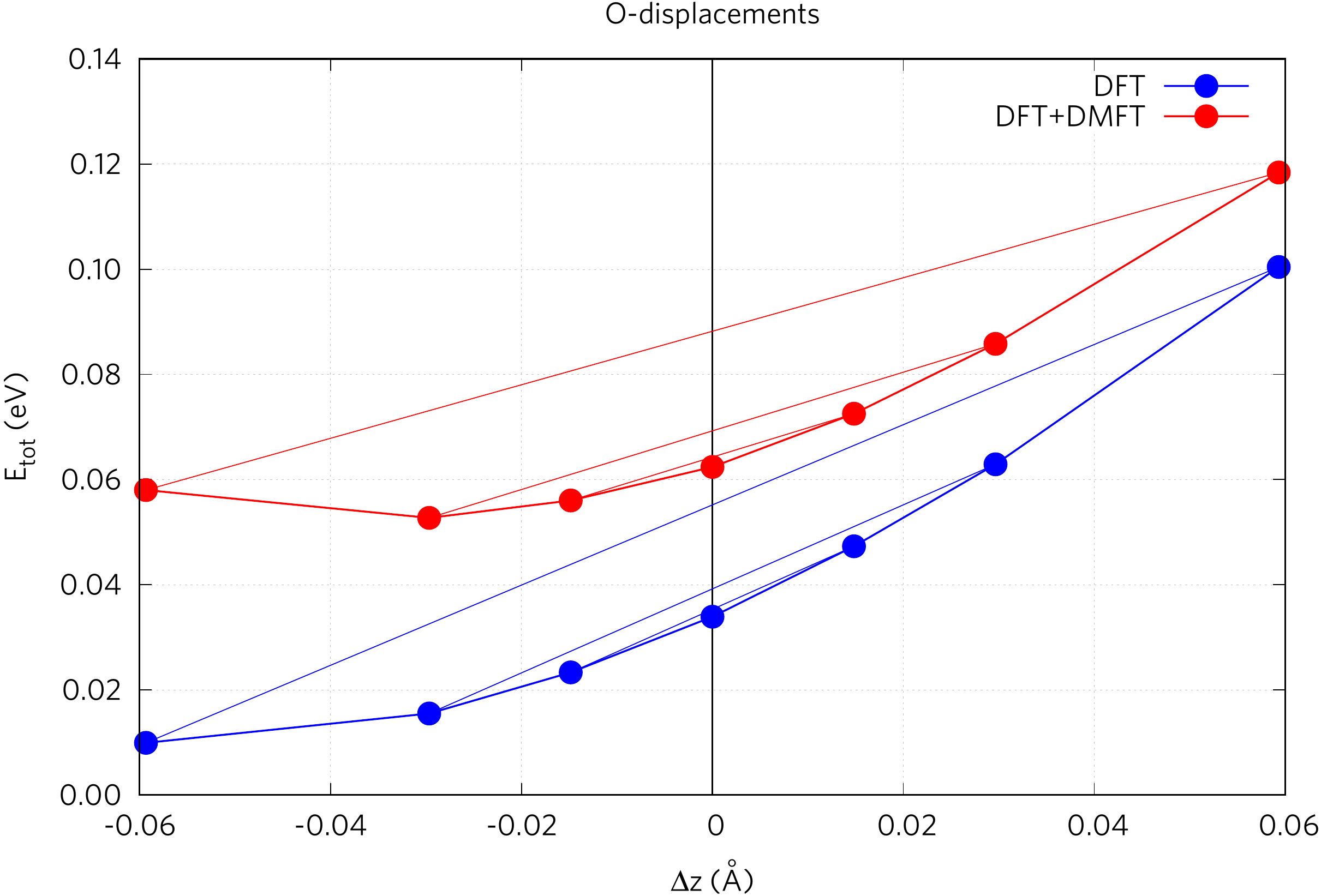}

\caption{\label{fig:Forces_parab}(Color online). The total energy as a function
of the $z$-position increments $\Delta z$ of Ce (upper panel) and
oxygen (lower panel) for three different increments: $\Delta z=4\%,\;2\%,\;1\%$
in units of $c$-axis lattice spacing. $a$ was kept equal to $3.81$\angstrom.
The energies are shifted in order to fit the graph.}
\end{figure}

\begin{figure}
\includegraphics[width=0.5\columnwidth]{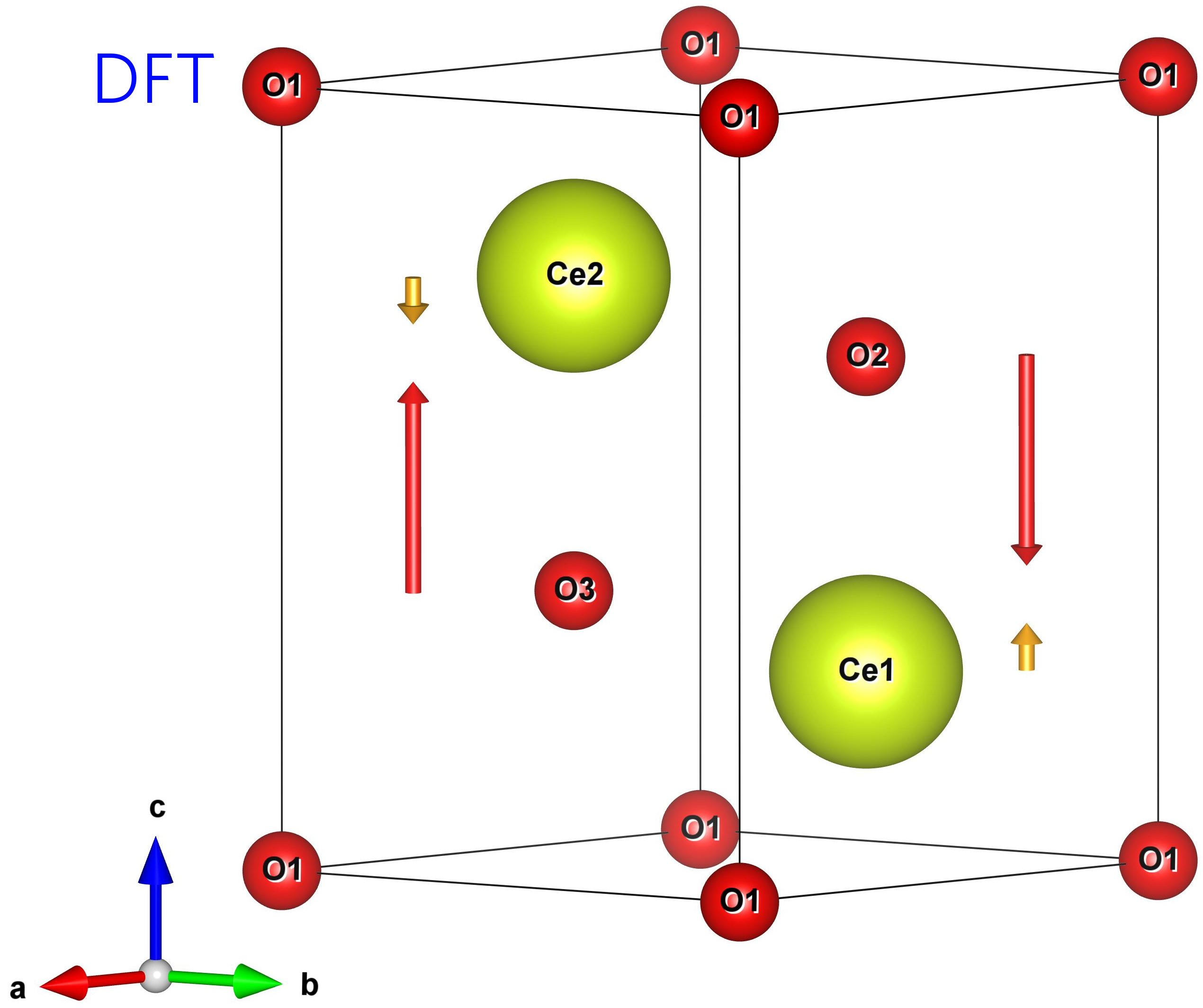}\includegraphics[width=0.5\columnwidth]{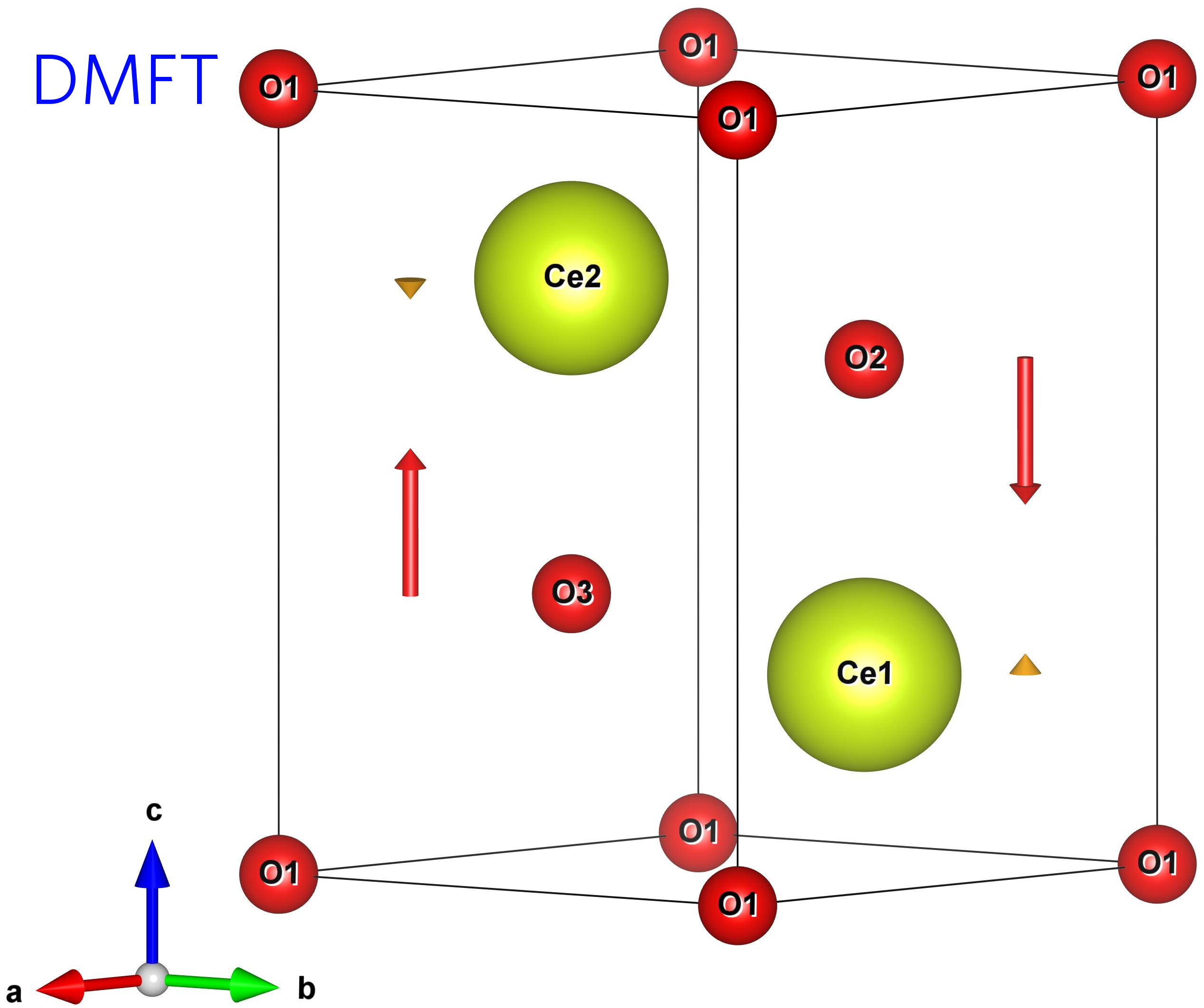}

\caption{\label{fig:Forces_im}(Color online). A graphical comparison of forces
calculated at the lattice constant $a=3.81\angstrom$ within DFT (left
panel) and DFT+DMFT (right panel). Forces acting on Ce$_{1(2)}$ and
O$_{2(3)}$ atoms are shown. The lengths of arrow are proportional
to the forces. Notice much smaller forces in case of DMFT.}
\end{figure}

In order to understand better the discrepancy between DFT+DMFT and
the experimental lattice constants in Ce$_{2}$O$_{3}$, we proceed
to calculate the atomic forces. For that purpose, we first note that
most internal atomic coordinates are fixed by symmetry. We vary the
remaining coordinates, which are the $z$-coordinates of Ce $2d$
and O $2d$ atoms (the ones established from experiment). Obviously,
the forces of the atoms related by symmetry are in turn related. During
finite increment of relevant atomic coordinates, we tested several
$\Delta z$ values, in order to be sure that the total energy varies
linearly over the lengthscale of $\Delta z$. The results of these
tests are shown in Fig.\ref{fig:Forces_parab}, where we report the
total energy profile for three different values of $\Delta z:\;4\%,\;2\%,\;1\%$
in units of the $c$-dimension of the unit cell. To ease the comparison,
we added thin lines, whose slope indicates the forces (up to the minus
sign):
\[
F_{z_{i}}=-\frac{\partial E_{tot}}{\partial z_{i}}.
\]
It can be seen from Fig.\ref{fig:Forces_parab}, that the slope remains
almost independent of $\Delta z$, therefore, in the following we
use $\Delta z=1\%.$ Table \ref{tab:Forces} summarises our results
for the atomic force calculations of Ce$_{2}$O$_{3}$. In addition,
we emphasise that the total energy as a function of $\Delta z$ is
a smooth differentiable function, thanks to the fact that both DFT
(CASTEP) and DMFT subsystems in our calculations are well-behaved,
giving small responses to small perturbations. Moreover, CASTEP DFT,
being a plane-wave code, does not introduce Pulay forces. We have
performed calculations for two lattice constants $a=3.81$\angstrom$\;$(minimum
energy for DFT+DMFT method) and $a=3.89$\angstrom$\;$(the experimental
value), while the ratio $c/a$ was kept fixed at the experimental
value $c/a=1.557$. We notice that taking into account strong correlations
of Ce $f$-shells within DMFT shows a systematic decrease of the forces
with respect to DFT, as illustrated in Fig.~\ref{fig:Forces_im}. This is the consequence of stronger cerium $f$-electron
charge localisation predicted by DMFT as compared to DFT, so that
these electrons participate less in formation of covalent bonds with
oxygen. This argument remains valid even though in our calculations
the electronic density is fixed: the total energy will be lower at
larger volumes in DMFT.

\begin{table}
\begin{tabular}{|c|c|c||c|c|}
\hline 
 & \multicolumn{2}{c||}{DFT} & \multicolumn{2}{c|}{DFT+DMFT}\tabularnewline
\hline 
 & $a=3.81$\angstrom & $a=3.89$\angstrom & $a=3.81$\angstrom & $a=3.89$\angstrom\tabularnewline
\hline 
\hline 
Ce & $\phantom{-}0.09$  & $\phantom{-}0.46$ & $-0.04$ & $\phantom{-}0.35$\tabularnewline
\hline 
O & $-0.40$ & $-0.30$  & $-0.28$ & $-0.17$\tabularnewline
\hline 
\end{tabular}

\caption{\label{tab:Forces} Atomic forces on Ce$_{2}$O$_{3}$'s Ce $2d$
and O $2d$ atoms, in units of $eV/\angstrom$. The forces are calculated
both with DFT and with DFT+DMFT at two values of the lattice constant:
the experimental value $a=3.89\angstrom$ and the value predicted
by DFT+DMFT $a=3.81\angstrom$.}
\end{table}

\section{\label{sec:Conclusions}Conclusions}

In conclusion, we have performed thorough DFT+DMFT studies of bulk
properties in $\gamma$-Ce, Ce$_{2}$O$_{3}$ and SmTe including bulk modulus,
equilibrium volume, forces and spectral weight.
We have used a newly
implemented DFT+DMFT formalism in the plane-wave code CASTEP, for
which case we re-derived all the necessary formulae. We have made
a comparison of our results with the ones available from literature.
The overall agreement of our spectral weight with the reference publication
of Ref.\onlinecite{Pourovskii_2007} is very good, taking into account
the difference in procedure employed to fix the chemical potential.
Our predicted bulk modulus somewhat better agrees with the experiment,
than what was previously published because of very well controlled underlying DFT
description given by CASTEP.
The same can be said about the equilibrium
volume (compared with the non-SC results of Ref.\onlinecite{Pourovskii_2007}):
our equilibrium volume for $\gamma$-Ce lays in between PAW/LDA+DMFT
and ASA/LDA+DMFT of Ref.\onlinecite{Amadon2012}, while for Ce$_{2}$O$_{3}$
our results are closer to ASA/LDA+DMFT.

The general effect
of applying DFT+DMFT on all the systems considered here is to increase the localisation of
the $f$-electrons with respect to DFT treatment, which
leads to larger estimates for equilibrium volumes.
In addition, we have also studied SmTe's equation of state and
demonstrated that improved treatment of strong localisation effects within
DFT+DMFT improves the agreement with the experiment as compared to LDA.
To the best of our knowledge, this is the first DFT+DMFT study of SmTe.
To understand why
DFT+DMFT does not fully recover the equilibrium volume in Ce$_{2}$O$_{3}$,
we performed the internal forces calculations corresponding to the
coordinates not constrained by symmetry. Our results show that within
the more realistic DFT+DMFT treatment, the atomic forces in cerium
sesquioxide appear to be profoundly different from the DFT ones.

A further point of novelty in our implementation is the ability to work
equally well both with norm-conserving and ultra-soft pseudopotentials,
since we accounted for the localised basis non-orthogonality via
introduction of scalar product metric. This approach could be of
interest when dealing with DMFT within {\textit e.g.} PAW formalism or
any other formalism, which uses non-orthogonal basis.

\begin{acknowledgments}
We wish to acknowledge support from EPSRC grant EP/M011038/1. We also
gratefully acknowledge the support of NVIDIA Corporation which donated
the Tesla K40 GPUs that were used for this research. In addition,
we are deeply thankful to M. Ferrero, O. Parcollet, L. Pourovskii
and A. Georges for fruitful discussion during the TRIQS developers'
meeting in Paris. This work used the ARCHER UK National Supercomputing
Service, for which access was obtained via the UKCP consortium and
funded by EPSRC grant ref EP/P022561/1.
\end{acknowledgments}

\appendix

\section{Details of Coulomb interaction treatment}

In the Hubbard-I solver, we use the most general form of Coulomb interaction
vertex (4-index):

\[
H_{ee}=\frac{1}{2}\sum U(m_{1},m_{2},m_{3},m_{4})c_{lm_{1},\sigma}^{\dagger}c_{lm_{2},\sigma^{\prime}}^{\dagger}c_{lm_{4},\sigma^{\prime}}^{\phantom{\dagger}}c_{lm_{3},\sigma}^{\phantom{\dagger}}.
\]
Internally, in the solver, in order to have a rough estimate of the
ground state sector, we have also used the reduced Coulomb interaction
vertex with two indices:

\begin{equation}
H_{ee}=\frac{1}{2}\sum U_{m_{1},m_{2}}^{\sigma,\sigma^{\prime}}n_{lm_{1},\sigma}n_{lm_{2},\sigma^{\prime}}.\label{eq:Coulomb2}
\end{equation}
Here the Coulomb matrix elements are expressed through $U(m_{1},m_{2},m_{3},m_{4})$
as follows:
\begin{align*}
U_{m_{1},m_{2}}^{\uparrow\uparrow} & =U_{m_{1},m_{2}}^{\downarrow\downarrow}=U(m_{1},m_{2},m_{1},m_{2})-U(m_{1},m_{2},m_{2},m_{1})\\
U_{m_{1},m_{2}}^{\uparrow\downarrow} & =U_{m_{1},m_{2}}^{\downarrow\uparrow}=U(m_{1},m_{2},m_{1},m_{2}).
\end{align*}
 Coulomb matrix elements $U(m_{1},m_{2},m_{3},m_{4})$ can be expressed
through the Slater integrals $F(n)$, assuming the spherical approximation\citep{Pourovskii_2007}:

\begin{widetext}

\begin{eqnarray}
U(m_{1},m_{2},m_{3},m_{4}) & = & \sum_{k=0}^{l}F(2k)\frac{4\pi}{2k+1}\sum_{q=-k}^{k}\langle lm_{1}|Y_{kq}|lm_{3}\rangle\langle lm_{2}|Y_{kq}^{\star}|lm_{4}\rangle=\label{eq:Coulomb4}\\
 & \sum_{k=0}^{l} & F(2k)(2l+1)^{2}\begin{pmatrix}l & k & l\\
0 & 0 & 0
\end{pmatrix}^{2}\sum_{q=-k}^{k}(-1)^{m_{1}+m_{2}+q}\begin{pmatrix}l & k & l\\
-m_{1} & q & m_{3}
\end{pmatrix}\begin{pmatrix}l & k & l\\
-m_{2} & -q & m_{4}
\end{pmatrix},\nonumber 
\end{eqnarray}

\end{widetext}

where $\begin{pmatrix}j_{1} & j_{2} & j_{3}\\
m_{1} & m_{2} & m_{3}
\end{pmatrix}$ are Wigner $3j$ symbols, while $Y_{kq}$ are spherical harmonics.
We report for completeness the relations among Slater integrals and
$U$ and $J$ for $d$- and $f$-orbitals. 

For $d$-orbitals with $l=2$, $2k$ in \eqref{eq:Coulomb4} runs
from $0$ to $4$ taking even values:

\begin{align*}
F(0) & =U\\
F(2) & =\frac{14J}{1.625}\\
F(4) & =0.625F(2).
\end{align*}

For $f$-orbitals with $l=3$ there is one more term $F(6)$, while
$F(2)$ and $F(4)$ are different respect to the previous case:

\begin{align*}
F(0) & =U\\
F(2) & =\frac{6435J}{286+\frac{195\times451}{675}+\frac{250\times1001}{2025}}\\
F(4) & =\frac{451}{675}J\\
F(6) & =\frac{1001}{2025}F(2).
\end{align*}

\section{\label{sec:app_Double}Double Counting Correction Schemes}

The double counting problem arises in both DFT+U and DFT+DMFT methods
since the amount of correlations present at the DFT level and originating
from the density functional is unknown. In order not to count the
same amount of correlations twice at both DFT and DMFT levels, we
need to adopt some model for DFT correlations and subtract this double
counting potential $V_{\sigma}^{DC}$ from the lattice Green function.
There are several approaches to this problem\citep{Anisimov_1991,Czyzyk_1994,Karolak_2010,Held_2007,Haule_2015,Park_2014}.
In CASTEP, we implement the following types of the double counting
corrections: i) Fully localized limit (FLL); ii) Around mean-field
limit (AMF)\citep{Anisimov_1991,Czyzyk_1994,Karolak_2010} and iii)
Held's mean-field one \citep{Held_2007}. The expressions for the
double-counting energy $E_{DC}$ and the double-counting potential
$V_{\sigma}^{DC}$ are reported below.
\begin{enumerate}
\item FLL: in this approximation, it is assumed that the occupation $n_{m\sigma}$
of an orbital $m,\sigma$ can be either $0$ or $1$. We denote $N_{\sigma}=\sum_{m}n_{m\sigma}$ and
$N_{tot}=\sum_{\sigma}N_{\sigma}.$ Then, from \eqref{eq:Coulomb2}
and assuming that $U_{m_{1},m_{2}}^{\sigma,\sigma^{\prime}}=U$ is
constant, we arrive at: 
\[
E_{DC}=\frac{1}{2}UN_{tot}\left(N_{tot}-1\right)-\frac{1}{2}J\sum_{\sigma}N_{\sigma}\left(N_{\sigma}-1\right).
\]
The double counting potential, can be obtained by differentiating
$E_{DC}$ with respect to $N_{\sigma}$ 
\[
V_{\sigma}^{DC}=U\left(N_{tot}-\frac{1}{2}\right)-J\left(N_{\sigma}-\frac{1}{2}\right).
\]
We note, that the above formulae remain valid also in the case when
$U_{m_{1},m_{2}}^{\sigma,\sigma^{\prime}}$ and $J$ are orbital dependent\citep{Czyzyk_1994}.
In that case, $U$ has the meaning of averaged Coulomb interaction.
It is assumed within FLL, that the electrons are fully localised,
hence it is normally suited to model insulating systems.
\item AMF: this is the opposite limit, where it is assumed that an average
occupation $n_{m\sigma}$ of an orbital $m,\sigma$ is independent
on $m$, so that 
\[
n_{m\sigma}=n_{\sigma}\equiv\frac{N_{\sigma}}{2l+1},
\]
where $N_{\sigma}$ is the total occupation of the impurity site in
the spin channel $\sigma$ and with $l$ orbitals. After some simplifications
we arrive at: 
\begin{align*}
E_{DC} & =UN_{\downarrow}N_{\uparrow}+\frac{2l}{2l+1}\frac{\left(U-J\right)}{2}\left(N_{\uparrow}^{2}+N_{\downarrow}^{2}\right)
\end{align*}
\[
V_{\sigma}^{DC}=U\left(N_{tot}-\frac{N_{\sigma}}{2l+1}\right)-JN_{\sigma}\left(\frac{2l}{2l+1}\right).
\]
This is somehow the opposite to FLL case and it is normally applied
to metals.
\item Held's formula: average Coulomb repulsion $\overline{U}$ is introduced
in order to ensure the rotational invariance as follows:
\[
\overline{U}=\frac{U+\left(l-1\right)\left(U-2J\right)+\left(l-1\right)\left(U-3J\right)}{2l-1}.
\]
Here $l$ is the degeneracy of the shell. The $E_{DC}$ and $V_{\sigma}^{DC}$
are then expressed as:
\begin{align*}
E_{DC} & =\frac{\overline{U}N_{tot}\left(N_{tot}-1\right)}{2}\\
V_{\sigma}^{DC} & =\overline{U}\left(N_{tot}-\frac{1}{2}\right).
\end{align*}
\end{enumerate}

\section{\label{sec:Matsubara_tails}Matsubara frequency summations}

We derive here an alternative form of Green function high-frequency
tails in Matsubara representation. We start by defining the spectral
moment expansion of the Green function up to $l$-th moment:

\begin{equation}
G(i\omega_{n})=\frac{a_{1}}{i\omega_{n}}+\frac{a_{2}}{(i\omega_{n})^{2}}+\frac{a_{3}}{(i\omega_{n})^{3}}+\ldots+\frac{a_{l}}{(i\omega_{n})^{l}}.\label{eq:G_moms}
\end{equation}
Here we assume $G$ and $\left\{ a_{i}\right\} $ to be matrices.
We assume that $\left\{ a_{i}\right\} $ are obtained \emph{e.g.}
by fitting the numerical data or by analytical calculations of Hamiltonian
commutators. As usual, we decompose the Green function into $G^{num}$
given by a numerical solution of the impurity problem and defined
up to a Matsubara frequency $\omega_{max}=\pi T(2n_{max}+1)$, and
$G^{an}(i\omega)=\sum_{m}\frac{a_{m}}{\left(i\omega\right)^{m}}$,
defined for all Matsubara frequencies. We then sum numerically 
\[
S_{1}=T\sum_{n=-n_{max}-1}^{n_{max}}\left(G^{num}(i\omega_{n})-G^{an}(i\omega_{n})\right)
\]
 and separately, analytically, $S_{2}=T\sum_{n}G^{an}(i\omega_{n})$.
The final result can be written as:
\[
T\sum_{n}G(i\omega_{n})e^{i\omega_{n}0^{+}}\approx S_{1}+S_{2}.
\]
We note that the sums here are extended over both positive and negative
Matsubara frequencies, and, hence, odd powers of $i\omega$ do not
contribute to $S_{2}$ (but must be included in $S_{1}$!). We report
below, the analytical formulae for even power contributions to $G^{an}$
up to $8$-th order. The coefficient $e^{i\omega_{n}0^{+}}$ is implied
in order to ensure the convergence: 

\begin{align}
T\sum_{i\omega_{n}}\frac{1}{i\omega_{n}} & =\frac{1}{2}\nonumber \\
T\sum_{i\omega_{n}}\frac{1}{(i\omega_{n})^{2}} & =-\frac{1}{4T}\nonumber \\
T\sum_{i\omega_{n}}\frac{1}{(i\omega_{n})^{4}} & =\frac{1}{48T^{3}}\label{eq:Tsums}\\
T\sum_{i\omega_{n}}\frac{1}{(i\omega_{n})^{6}} & =-\frac{1}{480T^{5}}\nonumber \\
T\sum_{i\omega_{n}}\frac{1}{(i\omega_{n})^{8}} & =\frac{17}{80640T^{7}}.\nonumber 
\end{align}
Calculation of the correlation energy within the Galitskii-Migdal
formula\eqref{eq:Migdal} can be, in principle, done in the same manner.
One only needs to express the tails of the product in terms of the
multipliers' tails. However, we find it more convenient to rewrite
the formula in another form, using the Dyson equation: $G_{0}^{-1}=G^{-1}+\Sigma$,
so that only the tails of $G$ are involved:

\begin{align*}
E_{corr} & =\frac{T}{2}\mathrm{Tr}\Sigma_{n}G(i\omega_{n})\Sigma(i\omega_{n})\\
\\
 & =\frac{T}{2}\mathrm{Tr}\Sigma_{n}\left(G_{0}^{-1}(i\omega_{n})G(i\omega_{n})-1\right).
\end{align*}
$G_{0}$ has a very simple form by construction. Moreover, $G_{0}^{-1}$
is at most a linear function of complex frequency:

\[
G_{0}^{-1}(i\omega_{n})=i\omega_{n}-\varepsilon_{0}.
\]
As above, we split $E_{corr}$ into $E_{num}$ and $E_{an}$:

\[
E_{corr}=E_{num}+E_{an},
\]
 where 

\[
E_{num}=\frac{T}{2}\mathrm{Tr}\sum_{n=-n_{max}-1}^{n_{max}}(i\omega_{n}-\varepsilon_{0})(G(i\omega_{n})-G^{an}(i\omega_{n}))
\]
and

\begin{equation}
E^{an}=\frac{T}{2}\sum_{\omega_{n}}\left\{ \sum_{l=1}^{N-1}\frac{\mathrm{Tr}(a_{l+1}-a_{l}\times\varepsilon_{0})}{(i\omega_{l})^{l}}+\frac{\mathrm{Tr(a_{N}\times\varepsilon_{0})}}{(i\omega_{n})^{N}}\right\} .\label{eq:Ean}
\end{equation}

where $G^{an}(i\omega_{n})$ is given by \eqref{eq:G_moms}. By using
formulae \eqref{eq:Tsums}, in \eqref{eq:Ean}, arrive at:
\begin{align*}
E^{an} & =\frac{1}{2}\left\{ \frac{1}{2}(\mathrm{Tr}(a_{2}-a_{1}\times\varepsilon_{0})-\frac{1}{4T}(\mathrm{Tr}(a_{3}-a_{2}\times\varepsilon_{0})\right.\\
 & +\left.\frac{1}{48T^{3}}(\mathrm{Tr}(a_{5}-a_{4}\times\varepsilon_{0})+\ldots\right\} .
\end{align*}
We remind, that here $\left\{ a_{i}\right\} $ and $\varepsilon_{0}$
are matrices, $\mathrm{Tr}$ is the usual trace operation on matrix,
while ``$\times$'' stands for matrix-matrix product. The advantage
of this method stays in the fact that we do not require the spectral
moments of the self-energy (which could be of worse quality), while
the expansion can be easily extended up to an arbitrary power of $i\omega$.
This calculation scheme is especially useful, when using Quantum Monte
Carlo solvers, in which there is an intrinsic bias in determination
of high-frequency tails. In our calculations, we used $l$ between
$5$ and $7$, which allowed to have a typical round-off error on
correlation energy around $10^{-6}$eV at a typical temperature of
$T=0.02eV$, as compared to an independently calculated value.

\bibliographystyle{apsrev4-1}
\bibliography{article_CASTEP}

merlin.mbs apsrev4-1.bst 2010-07-25 4.21a (PWD, AO, DPC) hacked
Control: key (0)
Control: author (72) initials jnrlst
Control: editor formatted (1) identically to author
Control: production of article title (-1) disabled
Control: page (0) single
Control: year (1) truncated
Control: production of eprint (0) enabled

\end{document}